\documentclass[prb,twocolumn,showpacs,preprintnumbers,amsmath,amssymb,floatfix,longbibliography]{revtex4-2}
\usepackage{graphicx}
\usepackage{float}
\usepackage{trfsigns}

\newcommand \be{\begin{eqnarray}}
\newcommand \ee{\end{eqnarray}}

\newcommand \ba{\begin{align}}
\newcommand \eea{\end{align}}

\newcommand \uu{\uparrow}

\DeclareMathOperator{\Tr}{Tr}

\begin{document}
           \csname @twocolumnfalse\endcsname
\title{Electronic quantum wires in extended quasiparticle picture}
\author{Klaus Morawetz$^{1,2}$,
 Vinod Ashokan$^3$, 
Kare Narain Pathak$^4$, Neil Drummond$^5$, Gianaurelio Cuniberti$^6$
}
\affiliation{$^1$M\"unster University of Applied Sciences,
Stegerwaldstrasse 39, 48565 Steinfurt, Germany}
\affiliation{$^2$International Institute of Physics- UFRN,
Campus Universit\'ario Lagoa nova,
59078-970 Natal, Brazil
}
\affiliation{$^3$Department of Physics, Dr. B. R. Ambedkar National Institute of Technology, Jalandhar 144011, India}
\affiliation{$^4$Centre for Advanced Study in Physics, Panjab University, 160014 Chandigarh, India}
\affiliation{$^5$Department of Physics, Lancaster University, Lancaster LA1 4YB, United Kingdom}
\affiliation{$^6$Institute for Materials Science,
TU Dresden,
01062 Dresden, Germany}

\begin{abstract}
Expanding the two-particle Green functions determines the selfenergy and the polarization as well as the response function on the same footing. 
The correlation energy is calculated with the help of the extended quasiparticle picture which accounts for off-shell effects. The corresponding response function leads to the same correlation energy as the selfenergy in agreement with perturbation theory provided one works in the extended quasiparticle picture. 
A one-dimensional quantum wire of Fermions is considered and ground state properties are calculated in the high density regime within the extended quasiparticle picture and Born approximation. While the on-shell selfenergies are strictly zero due to Pauli-blocking of elastic scattering, the off-shell behaviour shows a rich structure of a gap in the damping of excitation which is closed when the momentum approaches the Fermi one. The consistent spectral function is presented completing the first two energy-weighted sum rules. The excitation spectrum shows a splitting due to holons and antiholons as non-Fermi liquid behaviour. A renormalization procedure is proposed by subtracting an energy constant to render the Fock exchange energy finite. The effective mass derived from meanfield approximation shows a dip analogously to the onset of Peierls instability. The reduced density matrix or momentum distribution is calculated with the help of a Pad\'e regularization repairing deficiencies of the perturbation theory. A seemingly finite step at the Fermi energy indicating Fermi-liquid behaviour is repaired in this way.  
\end{abstract}
\maketitle

\section{Introduction and motivation}
The one-dimensional correlated electron gas is especially interesting since the quasiparticle picture breaks down and non-Fermi liquid behaviour appears. Such non-Fermi liquid behaviour has been observed in various physical systems ranging from large-scale structures like crystalline ion beams \cite{SSH01,SSBH02}, quantum wires \cite{PMC14}, carbon nanotubes \cite{Saito98,Bockrath99,Ishii03,Shiraishi03}, edge states in quantum hall liquid \cite{Milliken96,Mandal01,Chang03}, semiconducting nanowires \cite{Schafer08,Huang01}, cold atomic gases \cite{Monien98,Recati03,Moritz05} up to conducting molecules \cite{Nitzan03}. Mostly it is claimed that perturbation theory breaks down due to divergences in the expansion and the absence of quasiparticles since single-particle excitations turn into collective ones \cite{SDC19,DBG10,GVM93}. Nevertheless such excitations can show up eventually at the Luttinger surface where the Green functions have a zero at zero energy \cite{FA22}. Due to the absence of quasiparticles, expansions are necessary beyond the quasiparticle pole approximation. This is achieved if one expands with respect to small damping (scattering rate) resulting into the extended quasiparticle approximation \cite{SL94,SL95} and used for transport in impurity systems \cite{SLMa96,SLMb96} or nonlocal kinetic theory \cite{SLM96,MLS00,M17b}. 

The limit of small scattering rates was first 
introduced by \cite{C66a} for highly degenerated Fermi liquids, 
later used in \cite{SZ79,KKL84} for equilibrium nonideal plasmas. The 
same approximation, but under the name of the generalised
Beth-Uhlenbeck approach, has been used by \cite{SR87,MR94} in
nuclear matter studies of the correlated density or in the kinetic 
equation for nonideal gases \cite{BKKS96}. This extended quasiparticle picture is plagued by the same divergence at the Fermi energy for one-dimensional wires as it is typical for non-Fermi liquids. Recently this deficiency of quasiparticle picture has been cured by a Pad\'e approximation \cite{M23} which shows that the extended quasiparticle picture works and perturbation theory can be applied. This renewed interest in perturbation theory is motivated by the fact that in one-dimensional systems the strongly correlated case coincides with the small-density limit due to the special density dependencies of kinetic and correlation energy \cite{GPS08}. Therefore it is worth to investigate how far the extended quasiparticle picture which takes into account first-order damping and off-shell behaviour is able to capture the physics. 

The aim of the article is twofold. First in a pedagogical sense, different many-body schemes of Dyson equation and selfenergy on one side and the variational technique with response and correlation functions on the other side are unified and it is shown how they are yielding to identical results. New results are presented that half of correlation energy is stored in quasiparticle distribution compared to the Wigner function, the equivalent results of variational technique, charging formula and Dyson equation expansions of Green functions appear only within the extended quasiparticle picture. The second aim is to demonstrate the many-body scheme for a highly correlated system of one-dimensional Fermi wire. Here as new feature it is shown how a finite momentum distribution can be achieved with a proper Pad\'e regularization and perturbation theory can be applied. The effective mass signals a transition similar like Peierls transition. 

There exists of course a fast literature on many-body theories among them only Green function techniques are mentioned here \cite{T67,V94,G04,GV08,VMKA08,M17b}. Specially for one-dimensional systems exact solutions are known for Luttinger \cite{L63,LP74,ES07}, Tomonaga \cite{DL74}, and Gaudin-Yang models \cite{Ram17,Pan22} by the Bethe ansatz \cite{EFGKK10,GBML13}. Frequently bosonization techniques are used \cite{GGM10,Lu77,Solyom79,H99} due to the similar behaviour of long-distance correlations of Fermi and Bose systems \cite{H81a,Emery79}. In \cite{MVBP18} it was shown that the Random Phase Approximation (RPA) becomes exact in the high-density limit for one-dimensional systems. An overview about one-dimensional models can be found in \cite{V94,G04,GV08}. 

If one wants to consider more realistic systems like the width dependence \cite{GADMP22} of quantum wires one does not have any exact solutions and perturbation methods have been used to investigate analytically and numerically the ground-state properties \cite{LD11,LG16,Loos13}. Here the quantum Monte-Carlo method \cite{Casula2006, Shulenburger2008, LD11, Vinod18c} allows to simulate strongly coupled systems~\cite{Drummond2007} as e.g. implemented in the \textsc{casino} code \cite{Needs10}. Slater-Jastrow-backflow trial wave functions \cite{Drummond04,Lopez06} were used in these calculations. The simulation details can be found in \cite{Vinod18c}. The variational Monte-Carlo method system~\cite{LD11} and more accurate diffusion Monte Carlo can be treated as benchmark for the theory since it provides an exact solution for a well-defined model. In one dimension, diffusion Monte Carlo method is an exact method since the nodal surface is exactly known.

The paper consists of two main parts. First, in the second chapter two approaches to correlations are presented, i.e. the structure factor and the Dyson equation with selfenergy both rooted in the two-particle Green function. It will be shown that the extended quasiparticle picture reproduces the correlation results of structure factor and correlation energy. It is explicitly demonstrated on the Born approximation level. The coupling parameter integration as a special form of variational method is demonstrated to yield the same correlation energy. Second, in chapter III we present the model of finite-width Fermion wires and discuss systematically the Hartree-Fock and Born approximation. Within this paper we consider the Hartree-Fock as meanfield. From the meanfield the effective mass is calculated in chapter~\ref{effmass}. A dip occurs at twice the Fermi momentum indicating a similarity to the onset of Peierls instability. The selfenergy in Born approximation is then calculated in chapter~\ref{selfB} revealing an energy gap. The resulting spectral function is presented which requires a re-adaption of the correct pole when approaching selfconsistency. The extended quasiparticle approximation describes this correct poles and completes the first two energy-weighted sum rules. In chapter~\ref{SFCF} we collect the results of structure factor and pair correlation function together with the correlation energy. The reduced density matrix is explicitly calculated in chapter~\ref{reddd} which shows a divergence at the Fermi energy due to perturbation theory. With the help of a proper Pad\'e regulator this divergence can be subtracted and the momentum distribution takes a finite value at the Fermi momentum. The contact potentials as well as finite-size potentials show seemingly a finite jump at the Fermi energy like a Fermi liquid which is corrected by a proper Pad\'e regulator. Chapter~\ref{summ} summarizes the finding. In the appendices the corresponding integration schemes are presented for the selfenergy and the momentum distribution.

\section{Many-body scheme}

\subsection{Correlation energy and pair correlation\label{struct}}
We consider Hamiltonians of the form
\be
\hat H=\hat H_0+\hat V=\sum\limits_1  \epsilon_1\hat \Psi_1^+\hat \Psi_1+\frac 1 2 \sum\limits_{12} V_{12}\hat \Psi_1^+\hat \Psi_2^+\hat \Psi_2\hat \Psi_1
\label{0}
\ee
with creation operators $\Psi^+$ and the free single-particle band dispersion $\epsilon_1={k_1^2\over 2 m }$, e.g. for free particles. Numbers are cumulative variables, $1\equiv x_1,t_1,...$.
The probability to find a particle at $1$ and another at $2$ is expressed by 
the one-time pair-correlation function
\be
&&\langle \hat \Psi_1^+\hat \Psi_2^+\hat \Psi_2\hat \Psi_1\rangle_{t_1=t_2}=\langle \hat n\left (x_1\right )\hat n\left (x_2\right )\rangle -n(x_1)\delta(x_1-x_2)
\nonumber\\&&=
g(x_1,x_2)n\left (x_1\right )n\left (x_2\right ).
\label{gr}
\ee
This can be used to express the mean correlation energy in space representation by \cite{SL13}
\be
E_{int}&=&\frac 1 2 \int d x_1 dx_2 V(x_1,x_2)\langle \hat \Psi_{x_1}^+\hat \Psi_{x_2}^+\hat \Psi_{x_2}\hat \Psi_{x_1}\rangle 
\nonumber\\
&=&\frac 1 2 \int d x_1 dx_2 V(x_1,x_2)n(x_1)n(x_2)g(x_1,x_2)
\nonumber\\
&=&\frac 1 2 \int d x_1 dx_2 V(x_1,x_2)n(x_1)n(x_2)[g(x_1,x_2)-1]
\nonumber\\&&+\frac 1 2 \int d x_1 dx_2 V(x_1,x_2)n(x_1)n(x_2)
\ee
where the definition (\ref{gr}) is used in the second line and the Hartree energy $E_H$ is split off in the third line. Changing to difference and centre-of-mass coordinates $r=x_1-x_2$, $R=(x_1+x_2)/2$ and introducing the liquid structure function
\ba
n(R) S(r,R)=(g(r,R)\!-\!1) n\left (\!R\!+\!\frac r 2\right )\!n\!\left (\!R\!-\!\frac r 2\!\right )\!+\! n(R) \delta (\!r\!)
\label{2}
\end{align}
the interaction energy $E_{int}$ without Hartree term $E_H$ which is the correlation energy $E_c$ with Fock (exchange) term $E_F$, becomes
\be
E_{int}\!-\!E_H=E_c\!+\!E_F&=&\frac 1 2 \!\int\!\! d r dR\, V(r)n(R)[S(r,R)\!-\!1]
\nonumber\\
&=&\frac N 2 \int {d q\over (2\pi)^d} V_{-q}(S_q-1)
\label{ef}
\ee
where the last line is valid if $S(r,R)\approx S(r)$.
Integrating (\ref{2}) yields the density fluctuation correlator
\be
\int d r {\rm e}^{-irq} \int dR \,n(R) S(r,R)&=&\langle \hat n_q\hat n_{-q}\rangle-n_qn_{-q}\nonumber\\
&=&\langle\delta n_q\delta n_{-q}\rangle
\ee
where the left side can be integrated again for $S(r,R)\approx S(r)$ and finally
\be
S_q=\frac 1 N \langle\delta \hat n_q\delta \hat n_{-q}\rangle
\label{3}
\ee
with the total number of particles $N$.
It means that the structure function is the density fluctuation correlator.
Therefore it is advisable to investigate the density fluctuations.

\subsection{Two-particle Green's function and selfenergy}
\begin{figure}[h]
\centerline{
\includegraphics[width=7.5cm]{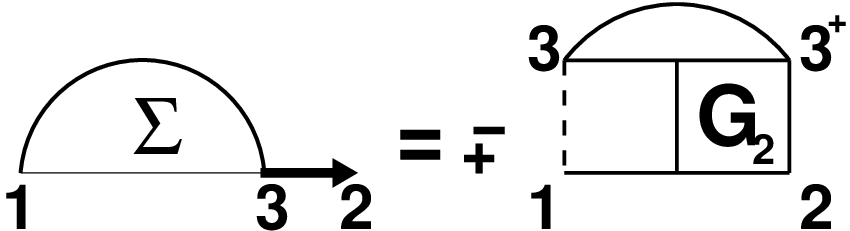}
}
\caption{Formal closure of the Martin-Schwinger hierarchy for Fermi/Bose systems introducing the selfenergy (\protect\ref{selfenergy}). About variable $3$ one integrates and $3^+$ means infinitesimal time later that at $3$. We introduce diagrammatic rules that a broken line means $i V$ and an arrow line means a causal Green's function $G$.}
\label{selfg2}
\end{figure}

For this aim we consider the causal Green's function
\be
G(1,2)=\frac 1 i \Theta(t_1\!-\!t_2) G^>(1,2)\!\mp\!
\frac 1 i \Theta (t_2\!-\!t_1) G^<(1,2)
\label{causalG}
\ee
for Fermi/Bose systems which time-orders the two double-time correlation functions
$
G^<(1,2)=
\langle \hat\psi^\dagger_2\hat\psi_1\rangle$ and
$G^>(1,2)=\langle \hat\psi_1 \hat\psi^\dagger_2\rangle$
with averaging about the unknown statistical operator. Applying the Heisenberg equation of motion one obtains the Martin-Schwinger hierarchy \cite{MS59,KB62} where the one-particle Green's function couples to the two-particle,
\be
G_2(1,3;2,4)={1\over i^2}\langle \hat T \hat \psi_1 \hat \psi_3 \hat \psi_{4}^\dagger\hat \psi_{2}^\dagger\rangle,
\label{G2}
\ee
the two-particle to the three-particle and so on. A formal closure is reached by introducing the selfenergy
\ba \label{selfenergy}
\mp i\!\int \!d3\, V(1,3) G_2(13,23^+)= \int\limits_{C}\! d3\,
\Sigma(1,{3}) G({3},2),
\end{align}
as illustrated in figure \ref{selfg2}. About double occurring indices we understand integration in the following. 
The integration path is determined by the demand that in the infinite past the two-particles are uncorrelated which leads to \cite{M17b}
\be
&&\int\limits_{C} d3 \Sigma (1,{3}) G(3,2) 
\nonumber\\
&&=
\int\limits_{t_0}^{+\infty} d{3} \biggl \{ \Sigma(1,3)
G(3,2) \mp \Sigma^< (1,3) G^> (3,2) \biggr \}.
\ee
This allows to set up conveniently the Langreth-Wilkins rules \cite{LW72} to recover the correlation parts from causal functions. If one has $C(1,2)=\int d3 A(1,3)B(3,2)$ these rules provide
\begin{equation}
C^\gtrless=A^\gtrless B^A+
A^RB^\gtrless
\label{bt24b}
\end{equation}
where the retarded/advanced functions are
\begin{eqnarray}
C^R(t,t^\prime)&=&-i\theta(t-t^\prime)\left[C^>(t,t^\prime)\pm C^<(t,t^\prime)\right],
\label{bt25}\nonumber\\
C^A(t,t^\prime)&=&\ \ i\theta(t^\prime-t)\left[C^>(t,t^\prime)\pm C^<(t,t^\prime)\right].
                   \label{bt26}
\end{eqnarray}

\begin{figure}[h]
\centerline{
\includegraphics[width=8cm]{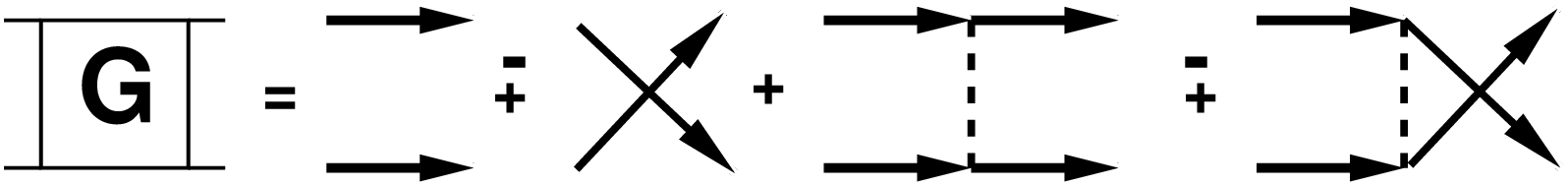}
}
\caption{Expansion of the two-particle Green function up to linear order in V.}
\label{g2_1}
\end{figure}

Now we can systematically expand the two-particle Green's function in terms of interaction as given in figure \ref{g2_1}. Introducing this into (\ref{selfenergy}), the first term gives the Hartree, the second the Fock, and the third and fourth term the first Born approximations of the selfenergy.

The closure (\ref{selfenergy}) leads to the Dyson equation for the full propagator
\be
G(1,2)=G_0(1,2)+G_0(1,3) \bar \Sigma(3,4)G(4,2)
\label{Dyson}
\ee
or graphically\\
\centerline{\includegraphics[width=8.4cm]{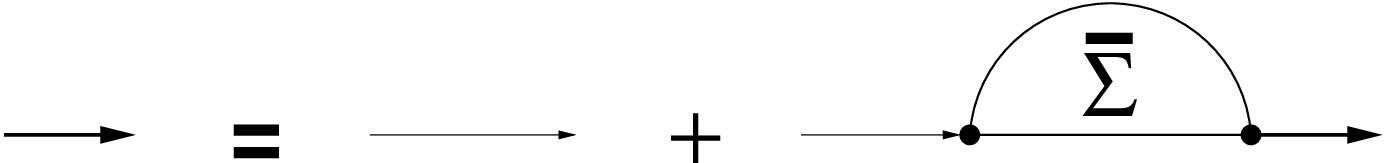}}
where the thin line is $G_0$ and we absorb the Hartree selfenergy together with the external potential $U$ into the induced potential
\be
{\bar U}_{11'}=U_{11'}\mp i V_{12} G_{22^+} \delta_{11'},
\label{6}
\ee
such that the free propagator reads
\be
\left (i\frac{\partial}{\partial
t_1}+\frac{\nabla_1^2}{2m}
-\bar U(1)\right ) G_0 (1,2) =
\delta(1-2).
\label{g0}
\ee
One Fourier transforms the difference coordinates into frequency/momentum. Since we concentrate on equilibrium all quantities are only dependent on the difference of coordinates. Nonequilibrium expressions can be found in \cite{M17b}.

\subsection{Correlation energy and extended quasiparticle picture}
From the Heisenberg equation of motion the Hamiltonian (\ref{0}) leads to
\ba
(i\partial_{t_1}\!-\!i\partial_{t_2})\Psi_2^+\Psi_1=&(\epsilon_1\!+\!\epsilon_2) \Psi_2^+\Psi_1\nonumber\\&+\sum\limits_3 (V_{13}+V_{12})\Psi_2^+\Psi_3^+\Psi_3\Psi_1
\end{align}
and therefore after averaging
\be
\left . \frac 1 2 (i\partial_{t_1}-i\partial_{t_2})G^<_{12}\right |_{1=2}=\langle K\rangle+2\langle V\rangle
\ee
with the kinetic energy $\langle K\rangle=\Tr [\hat \rho \hat H_0]$ and the potential energy  $\langle V\rangle=\Tr [\hat \rho \hat V]$.
If one knows the correlation function, the correlation energy density can be expressed therefore as
\be
E_c+E_{\rm H}+E_{\rm F}=\sum\limits_k \int {d\omega \over 2\pi} \frac 1 2 \left (\omega -{k^2\over 2 m}\right ) G^<(k,\omega)
\label{Ec}
\ee
where the total energy would be the plus sign instead of minus sign in the bracket. Here we note the different expansion scheme compared to (\ref{ef}) and the different role of Fock energy.

Within in the extended quasiparticle picture we expand the correlation functions with respect to the order of selfenergy
\ba
G^<(k,\omega)={2\pi\delta(\omega\!-\!\varepsilon_k)
\over 1-{\partial \Sigma(\omega)\over \partial \omega}}n_k\!+\!
\Sigma^<(\omega){\wp^\prime\over\omega\!-\!\varepsilon_k}+o(\Sigma^2)
\label{gs}
\end{align}
where the real part of the spectral function is the Hilbert transform 
\be 
\Sigma={\rm Re} \Sigma^R=\int {d\bar \omega \over 2 \pi}{\Gamma(\bar \omega)\over \omega-\bar \omega}
\label{Hilbert}
\ee
of the selfenergy spectral function
\be
\Gamma=\Sigma^> +\Sigma^<=-2 {\rm Im} \Sigma^R.
\label{Ga}
\ee
Both specifying the retarded selfenergy
\be
\Sigma^R(q,\omega)=\Sigma(q,\omega)-\frac i 2 \Gamma(q,\omega)=\int {d\bar \omega \over 2 \pi}{\Gamma(\bar \omega)\over \omega-\bar \omega+i\eta}.
\ee
We abbreviate
\be
\varepsilon^0_k=\epsilon_k+\Sigma^{\rm HF},\quad \varepsilon_k=\varepsilon^0_k+\Sigma(k,\varepsilon_k)
\ee
according to the poles of (\ref{gs}).

If we integrate (\ref{gs}) over the energy $\omega$ we get the connection between reduced density matrix $\rho$ and the distribution $n_k$ as
\be
\rho_k=n_k+\int {d\omega \over 2 \pi} {\Sigma^<(\omega)(1\mp n_k)-\Sigma^>(\bar \omega)n_k \over (\omega-\epsilon_k)^2}.
\label{red}
\ee
The quasiparticle (Bose/Fermi) distribution $n_k$ is to be take at the pole $\varepsilon_k$.
The form (\ref{gs}) was derived with respect to small scattering rate expansion and with quasiparticle energies under the name of extended quasiparticle approximation in nonequilibrium \cite{SL94,SL95} and the history of this expansion was given in the introduction above. Details can also be found in \cite{M17b}.

\subsection{Born approximation}
Since we want to consider the terms up to $V^2$ in the selfenergy, the terms exceeding (\ref{gs}) would start with fourth-order interaction. Using the expansion of figure \ref{g2_1} to calculate the selfenergy in figure \ref{selfg2} one sees that the Hartree and Fock terms are time-diagonal and therefore do not possess any frequency dependence which leads to no second part of (\ref{gs}). The Born terms lead to the selfenergies
\be
\Sigma^<(k,\omega)&=&\sum\limits_{q p} 2\pi \delta (\omega\!+\!\epsilon_p\!-\!\epsilon_{p\!-\!q}\!-\!\epsilon_{k\!+\!q}) n_{p\!-\!q} n_{k\!+\!q}(1\!\mp\!n_p)
\nonumber\\
&& \times V_q\left [g_s V_q\mp V_{p-k-q}\right ]
\label{Born}
\ee
where the spin degeneracy $g_s$ does only apply to the direct and not to the exchange terms. The $\Sigma^>$ selfenergy is obtained by interchanging $n\leftrightarrow 1\mp n$. 
Using this Born approximation in (\ref{red}) we get
\ba
&\langle \epsilon \rho\rangle=\sum\limits_k \epsilon_k \rho_k=\sum\limits_k \epsilon_k n_k
+\sum\limits_{kpq}V_q\left [g_s V_q\mp V_{p-k+q}\right ]\epsilon_k 
\nonumber\\&\times{n_{k+q}n_{p-q} (1\!\mp\! n_k)(1\mp n_{p})\!-\!n_k n_p (1 \!\mp\! n_{k+q})(1 \!\mp\! n_{p-q})\over (\epsilon_{k+q}+\epsilon_{p-q}-\epsilon_p-\epsilon_k)^2} 
\label{1h}
\end{align}
and
\be
\langle \rho\Sigma \rangle =\langle n \Sigma \rangle +o(V^4).
\label{2h}
\ee
Using symmetry to replace $\epsilon_k\to -\frac 1 4 (\epsilon_{k+q}+\epsilon_{p-q}-\epsilon_p-\epsilon_k)$ in (\ref{1h}) and subtracting (\ref{2h}) we obtain 
\ba
&\langle \varepsilon^0 \rho\rangle=\langle \varepsilon^0 n\rangle -\frac 1 4 \sum\limits_{kpq}V_q\left [g_s V_q\mp V_{p-k+q}\right ]
\nonumber\\&\times{n_{k+q}n_{p-q} (1\mp n_k)(1\mp n_{p})-n_k n_p (1 \mp n_{k+q})(1 \mp n_{p-q})\over \epsilon_{k+q}+\epsilon_{p-q}-\epsilon_p-\epsilon_k}.
\end{align}
We identify the difference of kinetic energy calculated with the reduced density matrix and the quasiparticle distribution function
\be
K_\rho&=K_n-\frac 1 2 E_c
\ee
with the 
first non-vanishing correlation energy (\ref{Ec})
\ba
{E_c}=&\int{dkdpd q\over (2\pi\hbar)^3}V_q[g_s V_q\mp V
_{p-k-q}]
\nonumber\\
&\times{
(1\mp n_{p-q})(1\mp n_{k+q}) n_{p}n_{k}\over 
\epsilon_{p}+\epsilon_{k}-\epsilon_{k+q}-\epsilon_{p-q}
}.
\label{ecorr}
\end{align}
The difference of reduced density matrix and quasiparticle distribution function accounts for half of correlation energy \cite{KM01} and we can write alternatively for the total energy
\be
E=K_n +\frac 1 2 E_c=K_\rho+E_c.
\ee 

We want to note here already that the on-shell selfenergy $\Sigma^\gtrless(k,\epsilon_k)$ will vanish for one-dimensional Fermi wires since two particles can scatter only by exchanging their momenta. In (\ref{gs}) the off-shell selfenergy is required which has been presented in a different scheme and discussed in \cite{MA23}.

Integrating (\ref{gs}) over frequency we get the reduced density matrix or Wigner distribution $\rho_k=n_k+\rho_k^>-\rho_k^<$ in terms of the Fermi-Dirac distribution $n_k$. We can explicitly using (\ref{Born}) in (\ref{red}) to obtain
\be
\rho_k^>&=&(1\mp n_k) \int{dpd q\over (2\pi\hbar)^2} 
{V_q(g_s V_q\mp V_{p-k-q})
\over 
(\epsilon_{p}+\epsilon_{k}-\epsilon_{k+q}-\epsilon_{p-q})^2
}
\nonumber\\
&&\times
n_{p-q}n_{k+q} (1\mp n_{p})
\label{rho}
\ee
and $\rho_k^<$ by replacing $n\leftrightarrow 1\mp n$.

\subsection{Density fluctuation and response function}
\begin{figure}[h]
\parbox[]{7.5cm}{
\includegraphics[width=7.5cm]{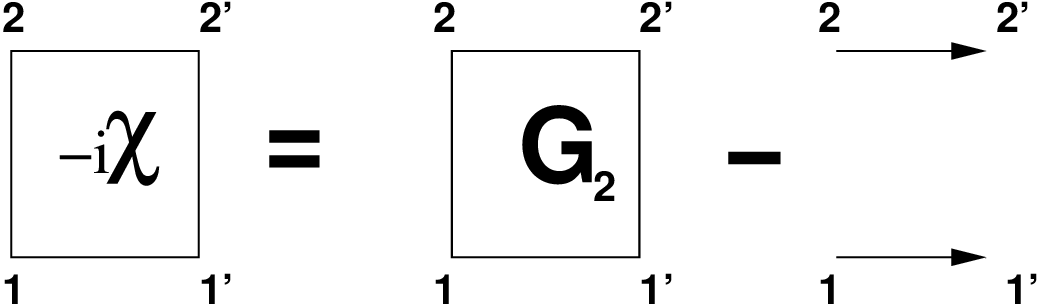}
}\hspace{0.3cm}
\parbox[t]{0.5cm}{
and
}
\hspace{0.3cm} 
\parbox[]{4cm}{
\includegraphics[width=4cm]{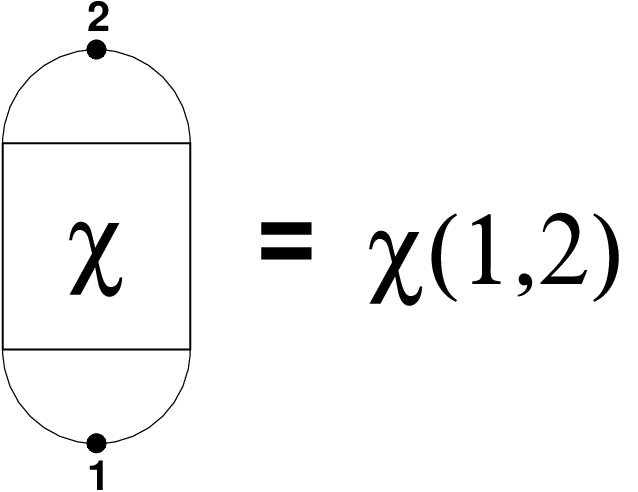}
}
\caption{The causal functions representing equation (\ref{nn}). \label{LG2}}\end{figure}

Besides the selfenergy we can extract also the density fluctuation $\delta \hat n (11^{\prime })=\Psi ^{+}(1^{\prime })\Psi (1)-\langle\Psi
^{+}(1^{\prime })\Psi (1)\rangle$ 
from the two-particle Green's functions. One observes that the fluctuation correlations are linked to the two-particle Green's function subtracted by the Hartree term
\begin{equation}
-i\chi(121^{\prime }2^{\prime })=G_2(121^{\prime }2^{\prime })-G(11^{\prime
})G(22^{\prime })  \label{defL}
\end{equation}
and 
\ba
\langle \delta \hat n (11)\delta \hat n (22)\rangle&=
\langle \Psi_1^+\Psi_1\Psi_2^+\Psi_2\rangle-G^<(11)G^<(22)
\nonumber\\
&=i^2\Theta(t_1-t_2) G_{121^+2^+}+G(11^+)G(22^+)
\nonumber\\&
=\chi^{>}(12)=\chi^{<}(21)
\label{nn}
\end{align}
which is illustrated in figure~\ref{LG2}.

In other words, instead of closing (\ref{selfenergy}) in an s-channel way, the density fluctuations consider the two-particle Green function in a t-channel, i.e. a different way to reduce the four times to two times. 
We can use now the causal two-time correlation function 
\begin{equation}
\chi(12)=\frac 1 i \Theta (t_1-t_2)\chi^{>}(12)+\frac 1 i \Theta (t_2-t_1)\chi^{<}(12)=\chi_{121^+2^+}
 \label{Ltwo}
\end{equation}
where we obey the Bose character of fluctuations by the plus sign in accordance with (\ref{causalG}).

On the other hand one can express the two-particle Green's function by a variation of the one-particle Green's functions $iG_{12}=\langle T_c\hat\Psi_1\hat\Psi_2^+ \rangle$ with respect to the external potential \cite{KB62,kker86}
\be
G_{121'2'}=G_{11'}G_{22'}\mp {\delta G_{11'}\over \delta U_{2'2}}
\label{1}
\ee
corresponding to Fermi/Bose systems. It shows that the response function is just the density fluctuation (\ref{Ltwo}) when we consider the times $t_2=t_1^+=t_1+0$
\be
\chi_{12}^<&=& {\delta n_1\over \delta U_{2^+2}}= {\delta G^<_{11^+}\over \delta U_{2^+2}}=\mp i {\delta G_{11^+}\over \delta U_{2^+2}}
\nonumber\\
&=&i (G_{121^+2^+}-G_{11^+}G_{22^+})=i \chi_{12}.
\label{chi}
\ee

Now we make the link to the structure function of chapter \ref{struct}.
From the definitions (\ref{gr})  and (\ref{2}) we see the relation of the pair correlation function and structure function to the fluctuation function (\ref{nn}) are
\be
i\chi(x_1t_1,x_2t_{1+})&=&\chi^<(12)=\langle \Psi_2^+\Psi_2\Psi_1^+ \Psi_1\rangle-n(1) n(2)
\nonumber\\
&=&\langle \Psi_1^+\Psi_2^+\Psi_2 \Psi_1\rangle+ \delta_{12} n(1)-n(1) n(2)
\nonumber\\
&=&n(1) n(2) [g(12)-1]+ \delta_{12} n(1)
\nonumber\\
&=&n(1)S(x_1,x_2,t,t)
\label{LLg}
\ee
which provides the static pair-correlation (\ref{gr}) and structure function (\ref{2}) in terms of the time-diagonal of $\chi^<$.  
Neglecting gradients in the density $n(R\pm r/2)\approx n(R)$
we obtain the known Fourier transform
\be
S_q&=&1+n\int d r {\rm e}^{-i q r} [g(r)-1]
={1\over n} \chi^>(q,t,t).
\label{structfac}
\ee
We can extend this to the dynamical structure factor by 
\be
n S(q,\omega)&=&
\int d(t-t'){\rm e}^{i\omega (t-t')}\chi^>(q,t,t')
\nonumber\\&=&\chi^>(q,\omega)=-2 {\rm Im} \chi(q,\omega) [1+n_B(\omega)]
\label{structd}
\ee
with the Bose distribution $n_B$ accounting for the Bose character of the fluctuations.
Since the imaginary part of the response function is odd in frequency we can write
\be
S_q&=&\frac 1 n \int\limits_{-\infty}^\infty{d\omega\over 2 \pi}[1+n_B(\omega)][-2 {\rm Im  }\chi(q,\omega)]
\nonumber\\
&=& \frac 1 n \int\limits_{0}^\infty{d\omega\over 2 \pi}{\rm coth}\left ({\beta \omega \over 2}\right )[-2 {\rm Im  }\chi(q,\omega)].
\label{cw}
\ee
If we consider the ground state at $\beta=1/k_BT\to \infty$ the $coth$ approaches unity.

\subsection{Coupling parameter integration}
So far we have two possibilities to calculate the correlation energy. We need the first-order expansion in the two-particle Green's function, Fig. \ref{g2_1}, in order to achieve the second order in the selfenergy in figure \ref{selfg2} and the correlation energy (\ref{Ec}). The same approximation of the two-particle Green's function is used now to calculate also the density response in figure \ref{LG2} as t-channel closing which provides the density fluctuations, the structure factor (\ref{cw}) and the correlation energy (\ref{ef}). Due to the same rooting of approximations to the two-particle Green's function the correlation energies (\ref{ef}) and (\ref{Ec}) will coincide.

The total ground-state energy can be obtained by coupling-constant integration. Therefore we add a constant $\lambda$ in front of the potential. Since the ground-state wave function is normalized independently of this parameter $\langle \Psi_0(\lambda)|\Psi_0(\lambda)\rangle=1$, the derivative of the ground-state energy reads
\ba
{d E(\lambda)\over d\lambda}&={d\over \lambda} \langle \Psi_0(\lambda)|\hat H(\lambda)|\Psi_0(\lambda)\rangle =\langle \Psi_0(\lambda)|{d\over \lambda}\hat H(\lambda)|\Psi_0(\lambda)\rangle \nonumber\\
&={E_{int}(\lambda)\over\lambda}
\end{align}
which provides the ground-state energy beyond the free one $E_0$ as
\be
E=E_0+\int\limits_0^1 {d\lambda\over \lambda} E_{int}(\lambda)
\label{charge}
\ee  
which we will call charging formula.
Using (\ref{cw}) in (\ref{ef}) we get the ground-state energy without Hartree term per particle as
\ba
{E-E_{H}\over N}=
-\frac{1}{2} \int \!{d q\over (2\pi)^d} V_{q} 
\left [\frac 1 n \int\limits_0^1 \!{d\lambda}\int\limits_{0}^\infty {d\omega\over\pi} {\rm Im}\chi(q,\omega,\lambda)\!+\!1\right ].
\label{ec}
\end{align}
Both forms (\ref{ec}) as well as (\ref{Ec}) allow a systematic expansion with the help of the two-particle Green's function. Some pitfalls of the coupling-constant integration are discussed in \cite{O76}.

\subsection{RPA-like integral equations}

\begin{figure}[h]
\centerline{\includegraphics[width=7cm]{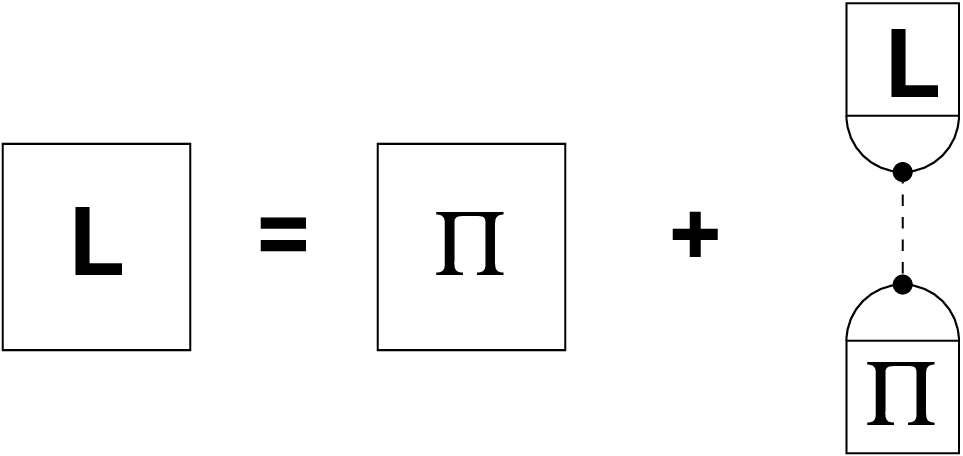}}
\caption{\label{v9} The integral equation connecting the response $\chi=iL$ with the polarization $\Pi$.}\end{figure}

Sometimes it is useful to express the response function (\ref{chi}) with respect to the external potential $U$ in terms of the polarization function which is the response with respect to the induced potential (\ref{6})
\be
\Pi_{12'1'2}=\mp {\delta G_{11'}\over \delta {\bar U}_{22'}}.
\ee
Using $\delta G=-G \delta G^{-1} G$ we can write the polarisation function with the help of the Dyson equation (\ref{Dyson})-(\ref{g0})
\be
\Pi_{12'1'2}=\mp G_{12}G_{2'1'}\mp G_{13}{\delta {\bar \Sigma}_{34}\over \delta {\bar U}_{22'}} G_{41'}.
\label{12}
\ee
Frequently one expresses the response function (\ref{chi}) or (\ref{defL}) in terms of the fluctuation $L$ with the help of (\ref{1}) as
\be
L_{121'2'}&=&G_{121'2'}-G_{11'}G_{22'}\nonumber\\
&=&\mp G_{12'}G_{21'}\mp G_{13}{\delta \Sigma_{34}\over \delta U_{2'2}} G_{41'}
\nonumber\\
&=&\mp G_{12'}G_{21'}+G_{13} {\delta \Sigma_{34}\over \delta G_{56}} L_{5262'}G_{41'}.
\label{L3}
\ee
Comparing (\ref{L3}) and (\ref{12}) and using the chain rule to express variations with respect to $U$ by variations with respect to ${\bar U}$ we find a relation between $L$ and $\Pi$ expressed in figure~\ref{v9}.

\begin{figure}
\centerline{\includegraphics[width=7cm]{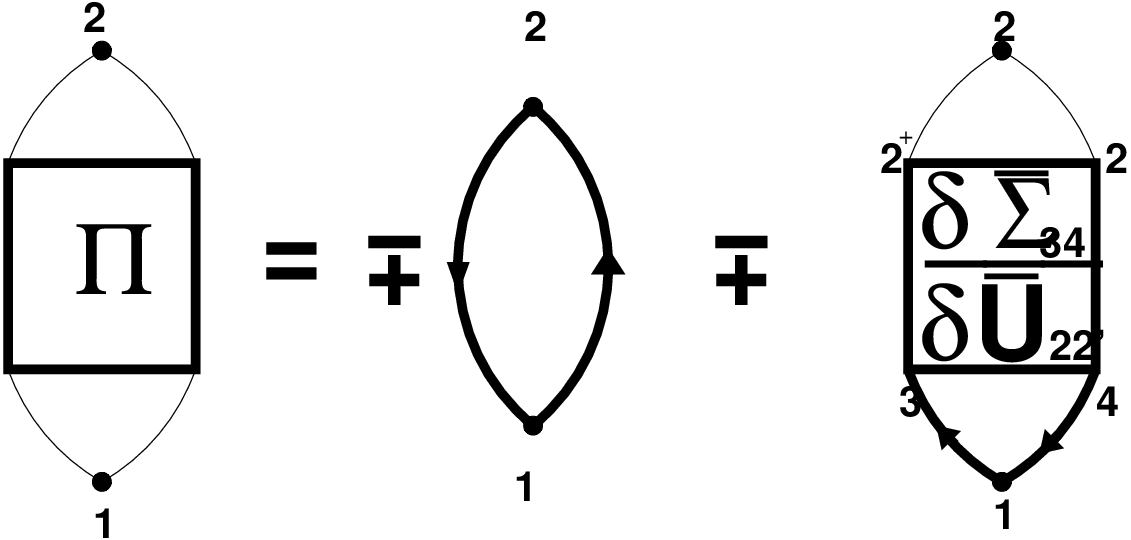}}
\caption{\label{v8a} Polarization (\ref{12}) in terms of the variation of the selfenergy with respect to the induced potential (\ref{6}). }
\end{figure}

Closing the upper and lower edges in the t-channel manner we obtain
the RPA-like integral equation for the causal functions
\be
L(12)=-i\chi(12)=\Pi(12)+\Pi(13)V(34)L(42)
\ee
which in equilibrium is solved and reads for the retarded functions
\be
L^R(q,\omega)=\chi^R(q,\omega)={\Pi^R(q,\omega)\over 1 -V(q) \Pi^R(q,\omega)}.
\label{LCHI}
\ee
There is some care to be observed since the polarization itself has a kernel to be determined selfconsistently with the selfenergy. In fact closing (\ref{12}) in the t-channel manner 
we get for the polarization just Fig. \ref{v8a} which shows that any approximation beyond the lowest RPA are calculated as variations of the selfenergy.

\begin{figure}
\centerline{\includegraphics[width=8cm]{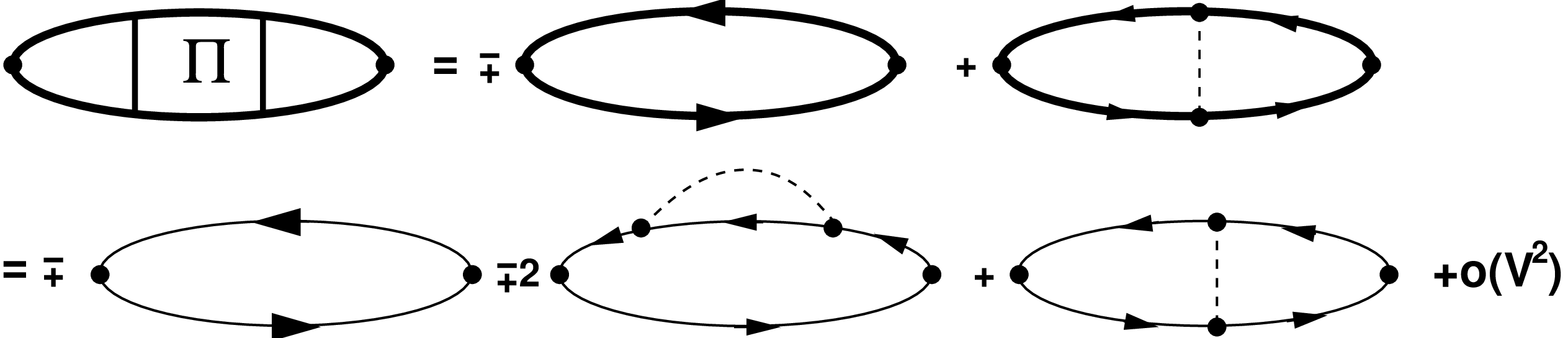}}
\caption{\label{pola} The polarization diagrams (\ref{12}) when introducing (\ref{var1}), the second line is the expansion with the help of the Dyson equation (\ref{Dyson}) up to first-order interaction $\mp \Pi_0\mp \Pi_{se}+\Pi_{ex}$. }
\end{figure}

Let us illustrate the procedure with the first-order expansion of the selfenergy which is the Fock term
\be
\Sigma^F(34)=\mp V(3-4)G(34)
\ee
and the Hartree term is absorbed in (\ref{6}).
We obtain
\ba
&{\delta \Sigma^F(3,4)\over \delta \bar U(2,2')}=\mp V(3,4) {\delta G(3,4)\over \delta \bar U(2,2')}
\nonumber\\&
=\pm V(3,4)G(3,5){\delta [\bar G_0^{-1}(5,6)+\Sigma(5,6)]\over \delta \bar U(2,2')} G(6,4)
\nonumber\\&
=\mp V(3,4)G(3,5)\delta_{2,6}\delta_{5,2'} G(6,4)+o(V^2)
\nonumber\\&
=\mp V(3,4)G(3,2')G(2,4)
\label{var1}
\end{align}
and introduced in (\ref{12}) one gets the expansion in figure \ref{pola}. Besides the non-interacting polarization function
\be
\Pi_{0}(q,\omega)=g_s\sum_{k} \frac{n_k-n_{k+q}}{\omega+\Omega_{k,q}}
\ee
there appear the selfenergy and vertex correction given by \cite{BMSP12}
\be
\Pi_{se}(q,\omega)=g_s\sum_{k,p}\frac{v(k-p)(n_k-n_{k+q})(n_p-n_{p+q})}{(\omega+\Omega_{k,q})^2}
\label{chi_se}
\ee
and
\be
\Pi_{ex}(q,\omega)=\mp g_s\!\sum_{k,p}\frac{v(k\!-\!p)(n_k\!-\!n_{k+q})(n_p\!-\!n_{p+q})}{(\omega+\Omega_{k,q})(\omega+\Omega_{p,q})},
\label{chi_ex}
\ee
respectively.
Here $\Omega_{k,q}=\omega_k-\omega_{k+q}$, $\Omega_{p,q}=\omega_p-\omega_{p+q}$ and $n_k$ represents the Fermi-Dirac or Bose distribution function for Fermions or Bosons respectively.

This completes the many body scheme where we have shown how the variational expressions resulting into pair correlation or structure function as well as response functions appear on the same footing as the single-particle self energy from Dyson equation. Both lead to the same expression for the correlated energy and are rooted to the two-particle Green function.

\section{Wire of Fermions}
\subsection{Model}
We apply now the many-body scheme to the model of a one-dimensional wire of charged Fermions.  Due to the strong divergence of the Coulomb interaction we model it by a soften Coulomb potential of a cylindrical wire $V(r)=e^2/4 \pi \epsilon_0 \sqrt{r^2+\bar b^2}$ and consider the limit $b\to 0$. Its Fourier transform reads
\ba
V(q)&={e^2\over 4 \pi \epsilon_0} v(q)
\nonumber\\
v(q)&=2 K_0(\bar b q)=
-2 \left[\ln \left(\frac{q}{2}\right)
+\gamma \right]-2 \ln \bar b
+o\left(\bar b^2\right)
\label{v}
\end{align}
where $\bar b$ is related to the transverse width parameter of the wire, $K_{0}$ is the modified Bessel function of $2^{nd}$ kind, and the Euler constant $\gamma$.

Within the jellium model of electron density $\rho(x)$ one considers an oppositely charged background density $\rho_b(x)$. The background potential $V_b(x)=-\int d x'V(x-x') \rho_b(x')$ gives the interaction energy of electrons with the background 
\be
E_{e-b}=-\int d x \rho(x) V_b(x).
\ee
This energy is compensated by the selfenergy of the background itself
\be
\frac 1 2 \int d x d x' \rho_b(x)\rho_b(x') V(x-x')
\ee
together with the Hartree selfenergy of the electrons
\be
\frac 1 2 \int d x d x' \rho(x)\rho(x') V(x-x')
\ee
if charge neutrality $\rho_b(x)=\rho(x)$ is assumed. Therefore the Hartree term does not count and we can use directly the formulas (\ref{Ec}) and (\ref{ec}) starting from the Fock term.

We consider first spin-polarized densities $n_{\uparrow\downarrow}=n(1\pm p)/2$ with arbitrary polarization 
$p=(n_\uparrow-n_\downarrow)/n$. Therefore the Fermi momentum is $k_{\uparrow\downarrow}=\pi\hbar n/g_s=k_F/g_s$ with $g_s=2/(1\pm p)$. For the paramagnetic case we have $p=0$ and $g_s=2$ which means $k_\uparrow=k_\downarrow=k_F/2$. Correspondingly for the ferromagnetic case $g_s=1$ and $k_\uparrow=k_F.$ The $r_s$ parameter as the number of particles in the Wigner size radius $2 a_B$ is $r_s=1/2n a_B$.

\subsection{Fock term or exchange term}

First we investigate the lowest order Fock term 
\be
\Sigma_F(k)=\mp \int\limits_{-\infty}^\infty {d q\over 2 \pi \hbar} V_q n_{k-q}
\label{F}
\ee
with the upper sign for spin-polarized electrons, $g_s=1$.
The spectral function in the Fock-propagator $G^<(k,\omega)=a(k,\omega) n_k$ becomes $a(k,\omega)=2\pi \delta(\omega-{k^2\over 2 m}-\Sigma_{\rm HF})$ and (\ref{Ec}) leads to the 
Fock correlation energy density
\be
{E_F\over \Omega}=\frac{1}{2}\int\limits_{-\infty}^\infty {d k\over 2 \pi \hbar} n_k \Sigma_F(k).
\label{fock1}
\ee
It is instructive to see how this formula appears from the charging formula (\ref{charge}) or (\ref{ec}).
For any temperature we have (\ref{structd}) and from (\ref{LCHI}) ${\rm Im}L={\rm Im}\chi={\rm Im}\Pi_0+o(V) $. With the help of $n_B(-\omega)=-1-n_B(\omega)$ we write
\ba
{E_{F}\over N}=
\frac{1}{2} \int \!{d q\over (2\pi)^d} V_{q} 
\left [\frac 1 n \int\limits_{-\infty}^\infty {d\omega\over\pi} n_b(-\omega){\rm Im}\Pi_0(q,\omega)\!-\!1\right ].
\label{ef1}
\end{align}
Using
\ba
{\rm Im} \Pi_0(q,\omega)=g_s \pi \!\int\!\! {d k\over (2\pi)^d} [n(\epsilon_k)\!-\!n(\epsilon_{k\!-\!q})] \delta (\epsilon_k\!-\!\epsilon_{k\!-\!q}\!-\!\omega)
\end{align}  
one calculates with the help of $n(a)[\pm 1-n(b)]=n(a)n(-b)=\pm n_B(a-b) [n(b)-n(a)]$
\ba
&\int\limits_{-\infty}^\infty {d\omega\over\pi} n_b(-\omega){\rm Im}\Pi_0(q,\omega)
\nonumber\\
&=g_s \int {d k\over (2\pi)^d} [n(\epsilon_k)-n(\epsilon_{k-q})] n_b(\epsilon_{k-q}-\epsilon_k)
\nonumber\\
&=\pm g_s \int {d k\over (2\pi)^d} n(\epsilon_{k-q}) n(-\epsilon_k)
\nonumber\\
&=g_s \int {d k\over (2\pi)^d} n(\epsilon_{k-q}) [1\mp n(\epsilon_k)]
\nonumber\\
&=n\mp g_s \int {d k\over (2\pi)^d} n(\epsilon_{k-q}) n(\epsilon_k)
\label{fockproof}
\end{align}  
and introducing into (\ref{ef1}) one gets exactly the Fock energy (\ref{fock1})
with (\ref{F}) 
\be
{E_F\over N}=\mp {g_s\over 2 n} \int {d k d q\over (2\pi)^{2d}}V_q n(\epsilon_{k-q}) n(\epsilon_k).
\label{fock2}
\ee

\begin{figure}[h]
\centerline{\includegraphics[width=8cm]{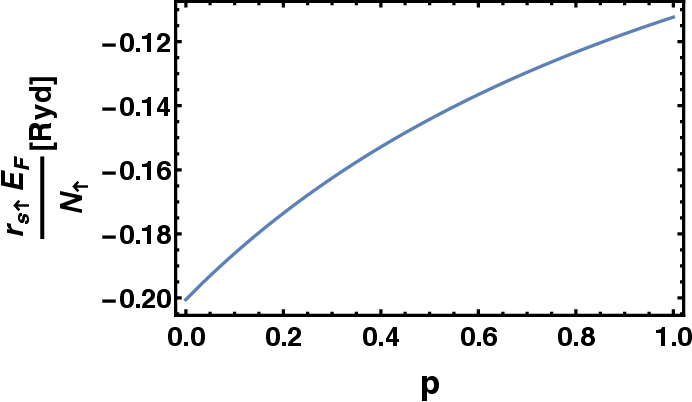}}
\caption{\label{fock} The Fock correlation energy for a width of $b=1a_B/\hbar$ versus polarization}
\end{figure}

We can give this exchange energy analytically at zero temperature
with $n_k=\Theta(k^2_{\uparrow \downarrow}-k^2)$.  In the following we scale the momenta in units of $k_{\uparrow\downarrow}=k_f(1\pm p)/2$ with spin-polarization $p=(n_\uparrow-n_\downarrow)/(n_\uparrow+n_\downarrow)$.
The Fock correlation energy per particle and in units of ${\rm Ryd}$ can be integrated with the potential (\ref{v}) by interchanging integration orders 
\ba
&
{E_F\over N_\uu {\rm Ryd}}={-1\over  16 r_{s\uu}} \!\!\int\limits_{-1}^1\!\! dk \!\!\!\int\limits_{k-1}^{k+1}\!\!\! d q K_0(b q)
={-1\over 8 r_{s\uu}}\!\! \int\limits_{0}^{2}\!\!\! d q (2\!-\!q) K_0(b q)
\nonumber\\&=-{1\over 8 r_{s\uu}}
\!\!\left [\!\frac{2 b (\pi  b \pmb{L}_0(2 b)\!+\!1) K_1(2 b)\!-\!1}{b^2}\!+\!2 \pi  \pmb{L}_{-1}(2 b)
   K_0(2 b)\!\right ]
\label{eFx}
\end{align}
with ${\rm Ryd}=e^2/4\pi \epsilon_0 a_B$ and $b=\bar b k_{\uparrow \downarrow}=\bar b k_F (1\pm p)/2$ and the StruveL function $L_n(x)$.
One sees that in this scaling the meanfield appears in orders $1/r_s$ or $r_{s\uu}=r_s/(1+p)$ respectively.
If one wants to present the energies in terms of Fermi energy one has the relation
\be
{\rm Ryd}={e^2\over 4\pi \epsilon_0a_B}=\epsilon_F {8\over \pi^2} r_s^2
\ee
and the mean field would start with $r_s$.

In figure~\ref{fock} the scaled Fock selfenergy per particle is plotted as a function of polarization. We see that it is increasing with increasing polarization.

\begin{figure}
\centerline{\includegraphics[width=8cm]{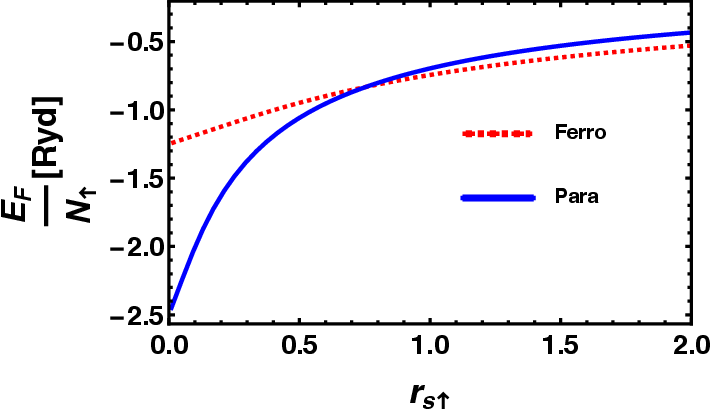}}
\caption{\label{ferro_para_b01} The ferro- and paramagnetic Fock correlation energy for a width of $b=0.1a_B/\hbar$}
\end{figure}

\begin{figure}
\centerline{\includegraphics[width=8cm]{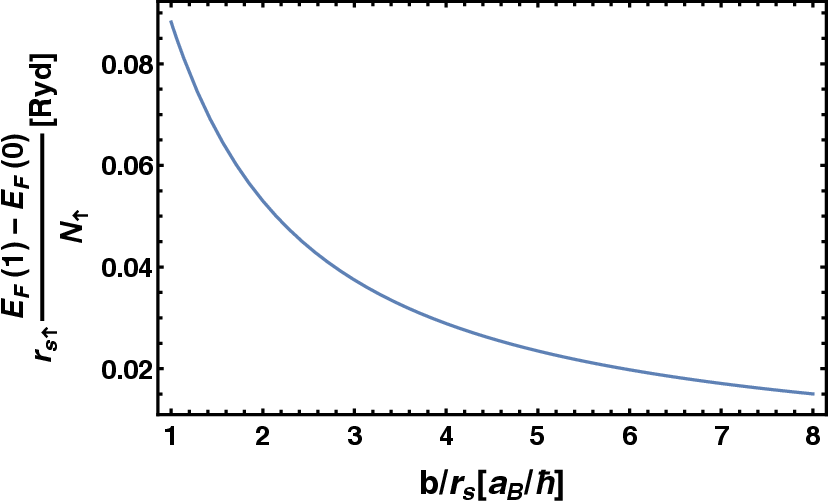}}
\caption{\label{ferro_para} The difference of ferro- and paramagnetic Fock correlation energy for any width.}
\end{figure}

Now we can investigate whether there is a symmetry-broken ground state by comparing the para- (p=0) with the ferromagnetic (p=1) ground state as illustrated in figure \ref{ferro_para_b01} and \ref{ferro_para}. We see that for any specific width the ferromagnetic ground state is higher than the paramagnetic one.  This is in agreement with the Lieb-Schultz-Mattis theorem \cite{LSM61} up to spin-up Bruckner parameter of $r_{s\uu}\sim0.7$ which shows the limit of meanfield approach.
If we scale the $b$ parameter we see that this is true for any width $b$ as illustrated in figure \ref{ferro_para}. We conclude that in one-dimensional systems there is no symmetry-broken Hartree-Fock state as found in 2D and 3D systems. For an overview see \cite{LG16}.

\subsection{Renormalization of potential and effective mass\label{effmass}}
\begin{figure}[h]
\centerline{\includegraphics[width=8cm]{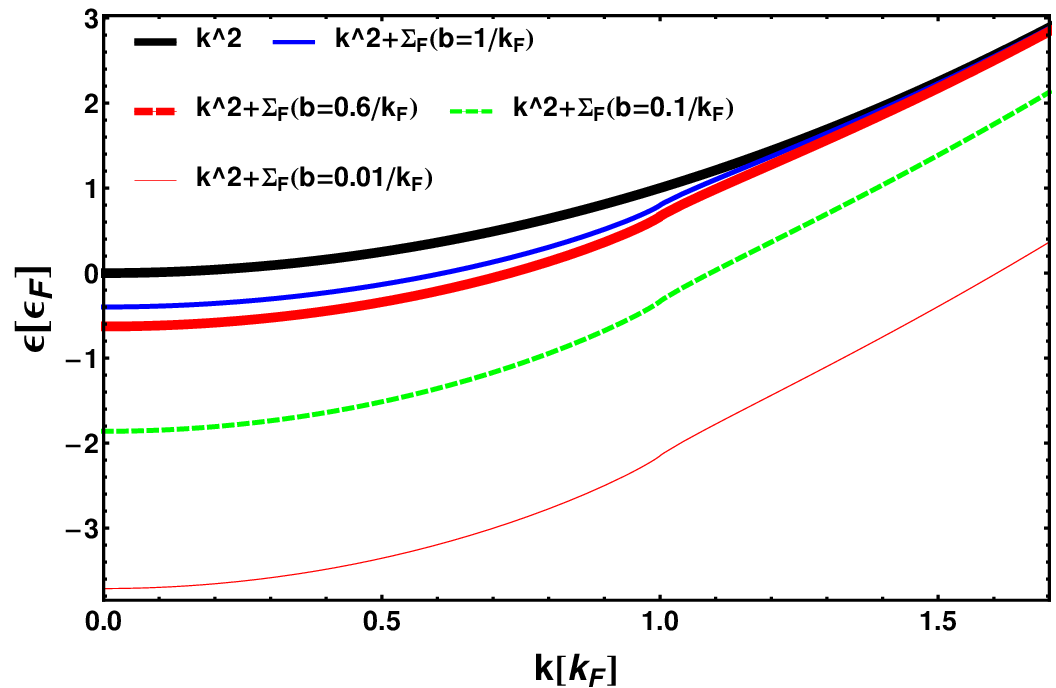}}
\caption{\label{HF} The un-renormalized Fock selfenergy (\ref{Fun}).}
\end{figure}

The analytical result for the Fock selfenergy is
\ba
{\Sigma_F\over n {\rm Ryd}}=&-{1\pm p\over 4 n r_s}\int\limits_{k-1}^{k+1} d q K_0(b q)
\nonumber\\
=&
-\frac{\pi}{4 r_s g_s}  \biggl [
(1-k) \pmb{L}_{-1}(b (1-k)) K_0(b \left| 1-k\right| )
\nonumber\\&
+\left| 1-k\right| 
   \pmb{L}_0(b (1-k)) K_1(b \left| 1-k\right| )
\nonumber\\&
+(k+1) \pmb{L}_{-1}(b (k+1)) K_0(b(k+1))
\nonumber\\&
+(k+1) \pmb{L}_0(b (k+1)) K_1(b (k+1))\biggr ]
\label{Fun}
\end{align}
which is plotted in figure \ref{HF}.

\begin{figure}[h]
\centerline{\includegraphics[width=8cm]{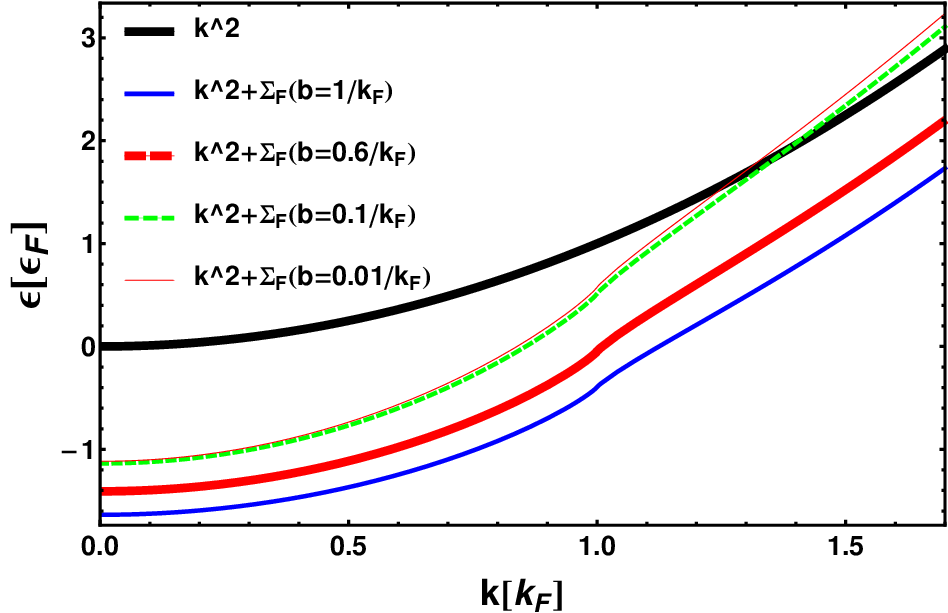}}
\caption{\label{HF_ren} The quasiparticle dispersion with the renormalized Fock selfenergy (\ref{Fun}).}
\end{figure}

When we are approaching the Coulomb limit for vanishing width $b\to 0$ the Fock term diverges to $-\infty$. This is cured by a summation of higher-order diagrams which are the RPA ones to produce an appropriate screened potential. Here we suggest the following procedure. In principal we can fix the energy scale as we want. Therefore adding a constant does not alter the physics. Within the jellium model we simply assume a background bias. Therefore we are allowed to subtract from any potential a constant, $v(q)\rightarrow v(q)+v_0$. Conveniently we chose $v_0=2 \ln b$ which renders the Fock selfenergy finite. We will see in the next chapter that this constant $v_0$ drops out exactly in second-order selfenergy. In figure \ref{HF_ren} we plot the renormalized Fock selfenergy as it appears in the quasiparticle dispersion
 which shows how it converges to a finite value for $b\to 0$, the result being
\ba
{\Sigma_F\over n {\rm Ryd}}=
\frac{\pi}{r_s g_s} \biggl [
\gamma\!-\!\ln{e^2\over  2}\!+\!{k\!+\!1\over 2}\ln(k\!+\!1)\!+\!{1\!-\!k\over 2}\ln|1\!-\!k|\biggr ].
\label{sfo}
\end{align}

The effective mass
\be
{k\over m_{\rm eff}}={\partial \varepsilon^0\over k}={k\over m}+{\partial \Sigma_F\over \partial k}
\ee
is independent of such added constant and diverges logarithmically at $k=1$
\be
{m\over m_{\rm eff}}&=&1\!-\!(p\!+\!1) \frac{r_s}{\pi
   ^2}
 \left[\ln \left(\frac{b}{2}|k\!-\!1|\right)+K_0(2 b)+\gamma \right]
\nonumber\\
&&+o(|k-1|)
\ee

\begin{figure}
\centerline{\includegraphics[width=8cm]{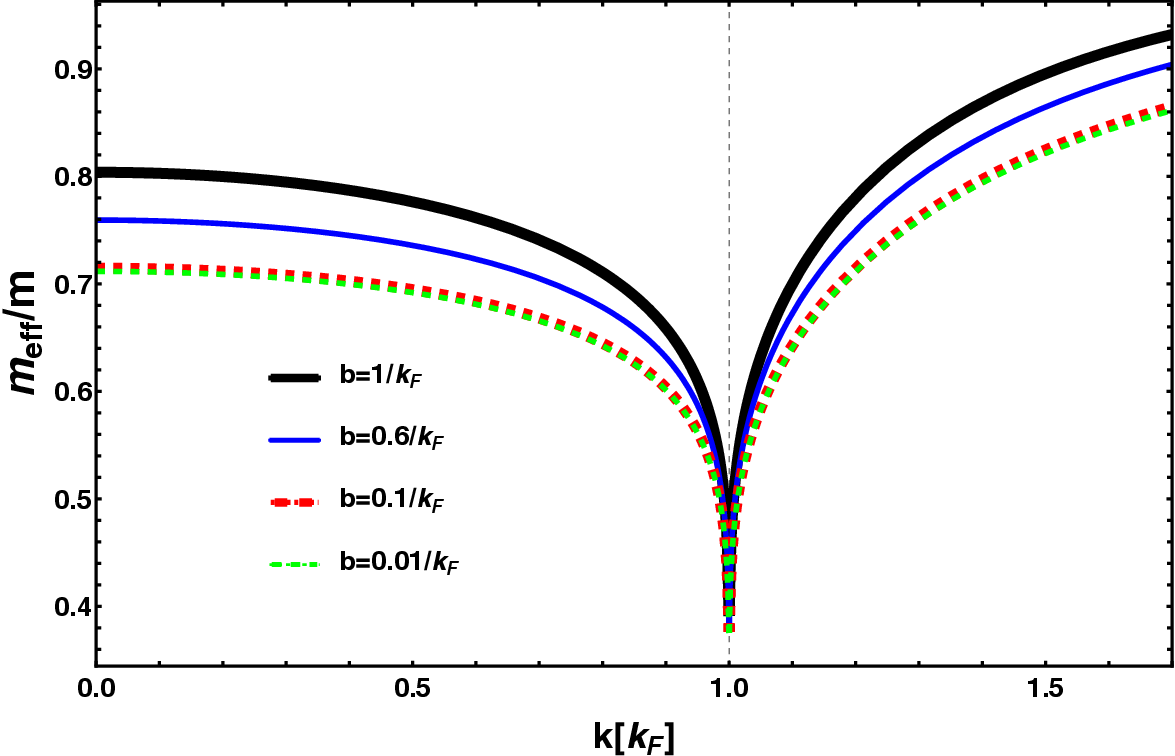}}
\caption{\label{meff} The effective mass for different width in the ferromagnetic case and $r_s=1$. The position of the logarithmical divergence are indicated by the dashed line.}
\end{figure}

In figure \ref{meff} we see that the Coulomb limit $b\to 0$ is reached with a finite value
\be
{m\over m_{\rm eff}}=1+r_s{2(1\pm p)\over x \pi^2}\ln{1+x\over |1-x|}
\ee
with $x=k/k_{\uparrow\downarrow}$.
At the (polarized) Fermi momentum we see that the effective mass is zero indicating the break down of the Fermi liquid picture.
 
\begin{figure}
\centerline{\includegraphics[width=8cm]{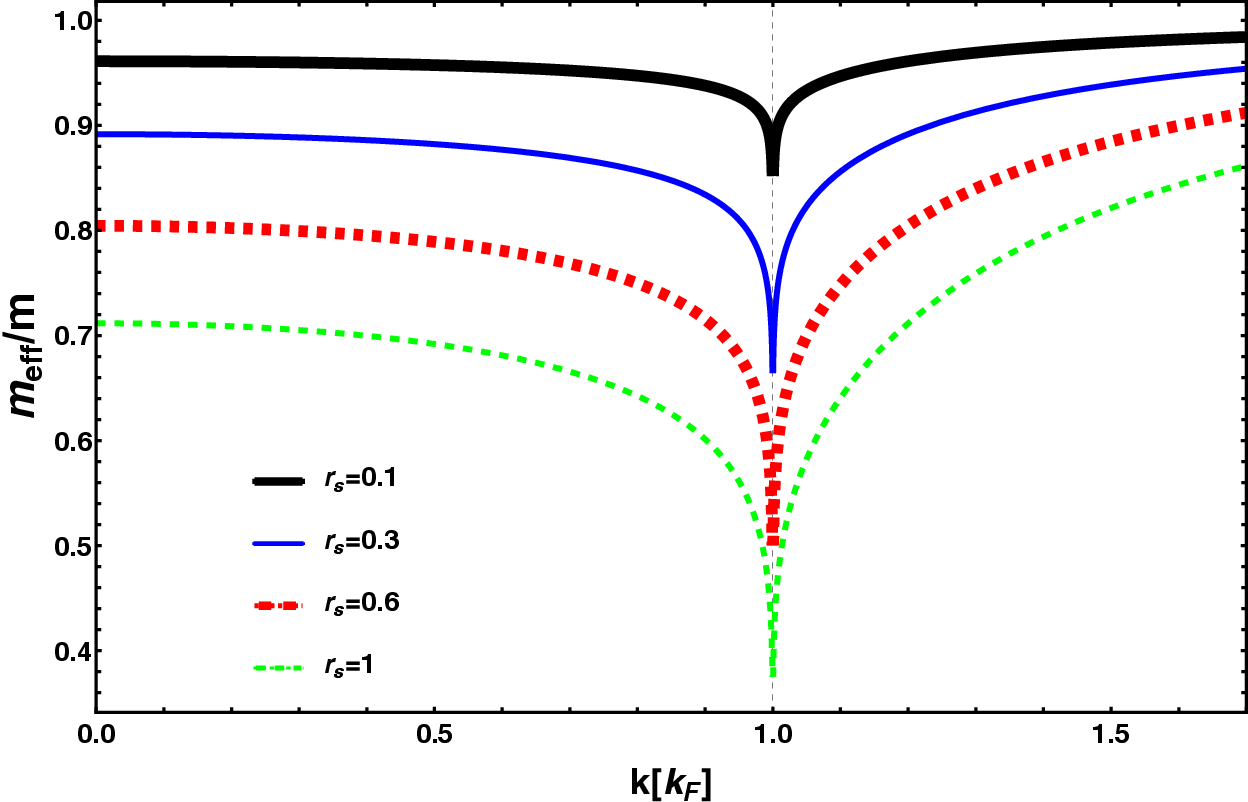}}
\caption{\label{meff_b01_p1} The Coulomb effective mass ($b=0.01$) for different $r_s$ in the ferromagnetic case.}
\end{figure}

\begin{figure}
\centerline{\includegraphics[width=8cm]{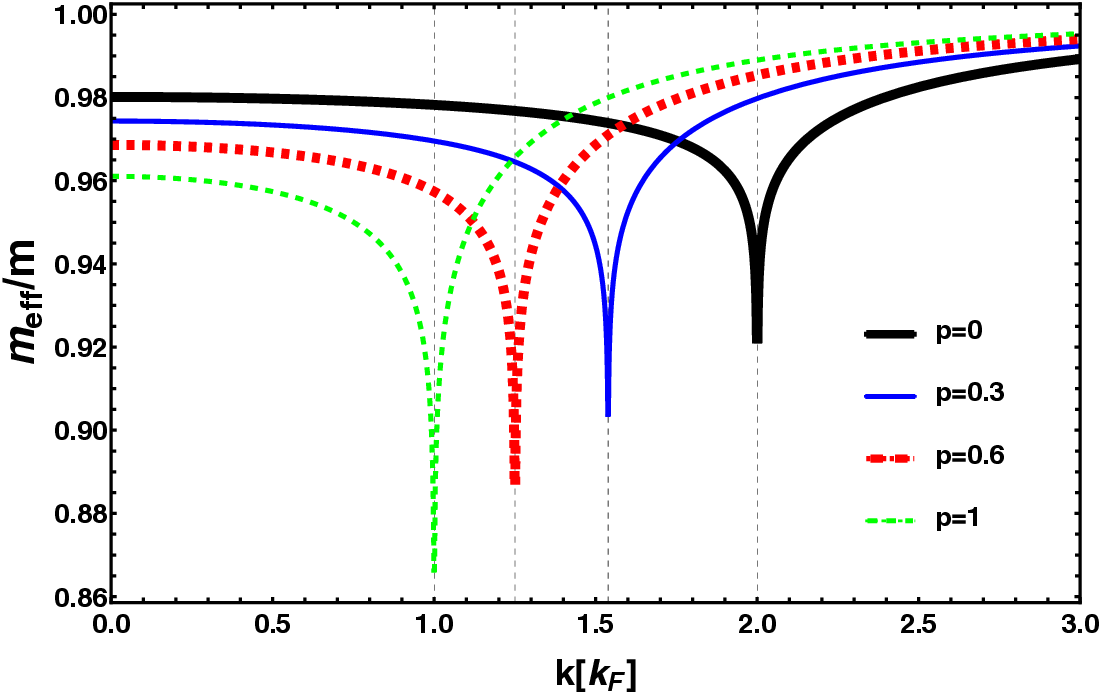}}
\caption{\label{meff_p} The Coulomb effective mass for $r_s=0.1$ and various polarizations ranging from paramagnetic (p=0) to the ferromagnetic (p=1) case. The vertical dashed lines indicate the divergence at $x=k/k_{\uparrow\downarrow}=2/(1+p)$}
\end{figure}

This Coulomb effective mass we plot in figure \ref{meff_b01_p1} for different $r_s$ parameter. 
We see that with increasing density the effective mass is more suppressed. The dip in the effective mass is dependent on the polarization as seen in figure \ref{meff_p}. In fact for the paramagnetic case we see that the dip occurs at twice the Fermi momentum and indicates an analogous onset of Peierls instability \cite{KE87,KoS96} though we do not have any lattice in the Hamiltonian. The formation of Wigner lattice due to correlation in experiments \cite{SSH01,SSBH02} allows to suggest here a similar transition.

\subsection{Selfenergy in first Born approximation\label{selfB}}

Now we calculate the selfenergy in Born approximation (\ref{Born}) which represents the next order in $r_S$ beyond the meanfield. The $\delta$-function we use to perform the $q$-integration which gives two poles $q=(p-k\pm \eta)/2$ with $\eta=\sqrt{p^2-k^2-2 k p +2 \omega}$ with the residue $1/2 \eta$. This restricts the integration to render the root real. The sum of this two poles yields finally
in dimensionless units
\be
&&{\Sigma^<\over {\rm Ryd}}={g_s\over \pi} \int {dp\over \eta} 
\left [v\left ({p-k+\eta\over 2}\right )-v\left ({p-k-\eta\over 2}\right )
\right ]^2 
\nonumber\\
&&\Theta[4-(k+p+\eta)^2]\Theta[4-(k+p-\eta)^2]\Theta[p^2-1].
\label{self1}
\ee
The expression for $\Sigma^>$ is given by inter-changing the sign in the $\Theta$ functions.
 It is remarkable that any constant shift of the potential $v(q)+v_0$ drops out.
Therefore we can work with the renormalized potential as introduced in the meanfield section. The last integration can be done numerically. The integration range for $p$ is in fact quite involved and given in appendix (\ref{range}). In \cite{MA23} an alternative analytical way is presented to express the selfenergy in terms of one integral about any used potential. 

The first observation is that both selfenergies vanish on-shell $\Sigma^\gtrless(k,\omega=k^2/2m)=0$. This is a specific feature of 1D systems. 
One can understand this as suppression of any elastic scattering event by Pauli-blocking allowing only exchange of momenta.

\begin{figure}
\centerline{\includegraphics[width=8cm]{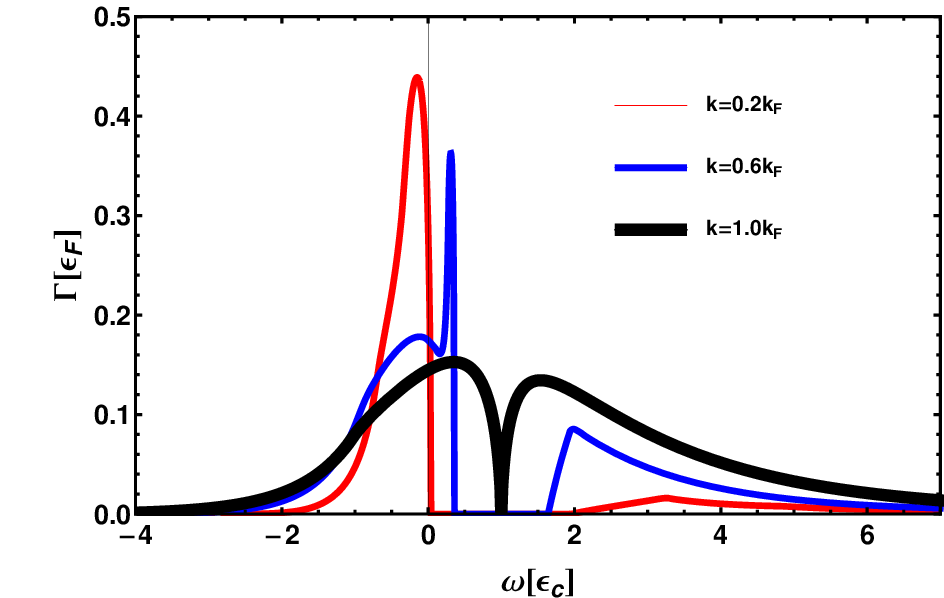}}
\centerline{\includegraphics[width=8cm]{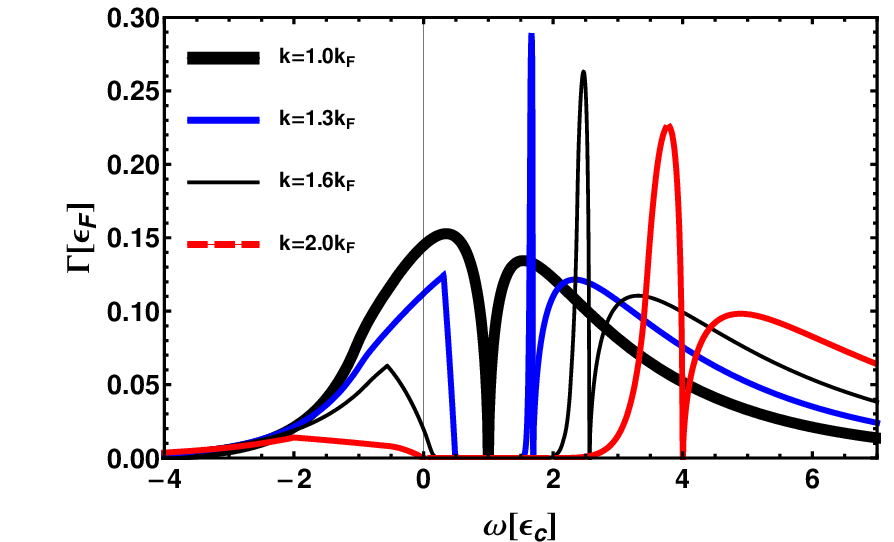}}
\caption{\label{gam} The spectral function of selfenergy (\ref{Ga}) for $b=1$ and various momenta. The left curves are $\Sigma^<$ and the right ones $\Sigma^>$.}
\end{figure}

\begin{figure}
\centerline{\includegraphics[width=8cm]{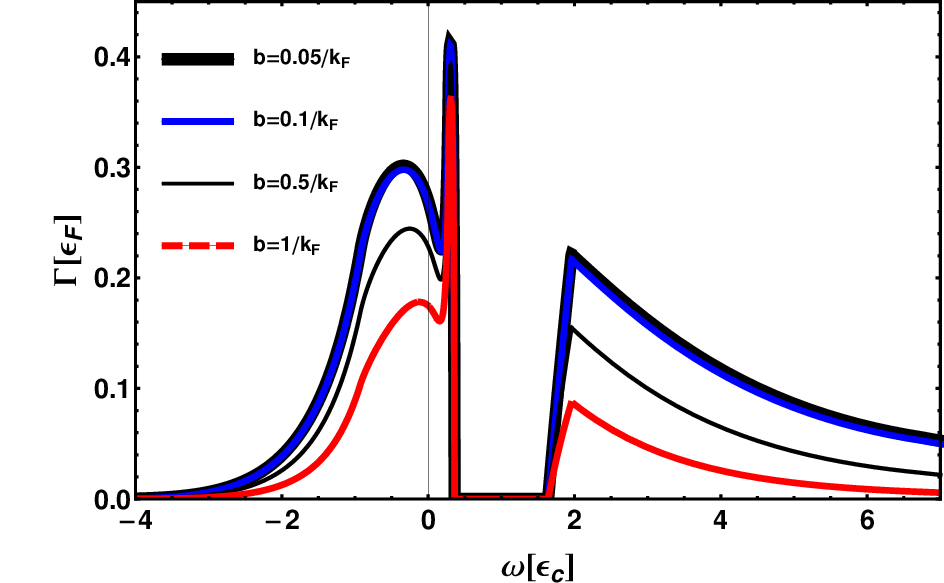}}
\centerline{\includegraphics[width=8cm]{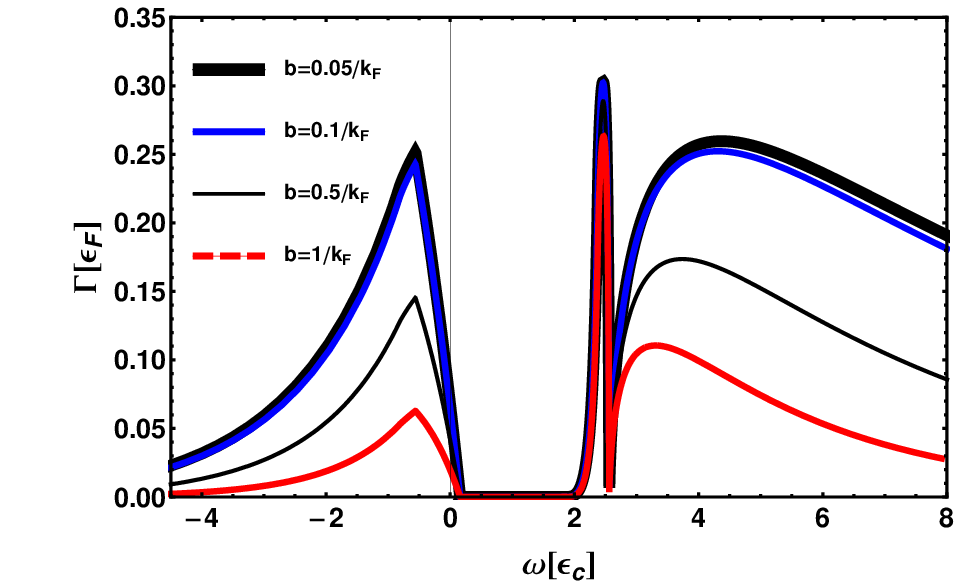}}
\caption{\label{gam2} The spectral function of selfenergy (\ref{Ga}) for $k=0.6$ (above) and $k=1.6$ (below) and various $b$.}
\end{figure}

We discuss the spectral function (\ref{Ga}) of the selfenergy in figure~\ref{gam}.
One sees that below the Fermi momentum a gap in the dissipation spectrum appears which is closed when the Fermi momentum is approached. For momenta above the Fermi momentum, the dispersion splits which can be interpreted as holons and antiholons \cite{EFGKK10}, i.e. the excitation out of Fermi see above $k_f$ and $-k_f$ respectively. This results into the two excitations above and below $\omega=k_f^2$. The $\Sigma^<$ as selfenergy due to hole damping represents the left curves and develop a sharp peak when approaching the Fermi momentum. It never overcomes the on-shell value. Exceeding the Fermi momentum $\Sigma^<$ shrinks and form a large background. The opposite behaviour one sees for the particle contribution $\Sigma^>$ which are the curves on the right side respectively. The sharp peak developed above the Fermi momentum is moving to higher values and broadens for higher momenta. Please note that at the on-shell value $\Sigma^>$ is also exactly zero. If we approach the Coulomb limit for $b\to 0$ we see in figure \ref{gam2} that the selfenergy is increasing and converging visibly at $b=0.05$.

\subsection{One-particle spectral function}
\subsubsection{Selfconsistent spectral function}
\begin{figure}
\centerline{\includegraphics[width=8cm]{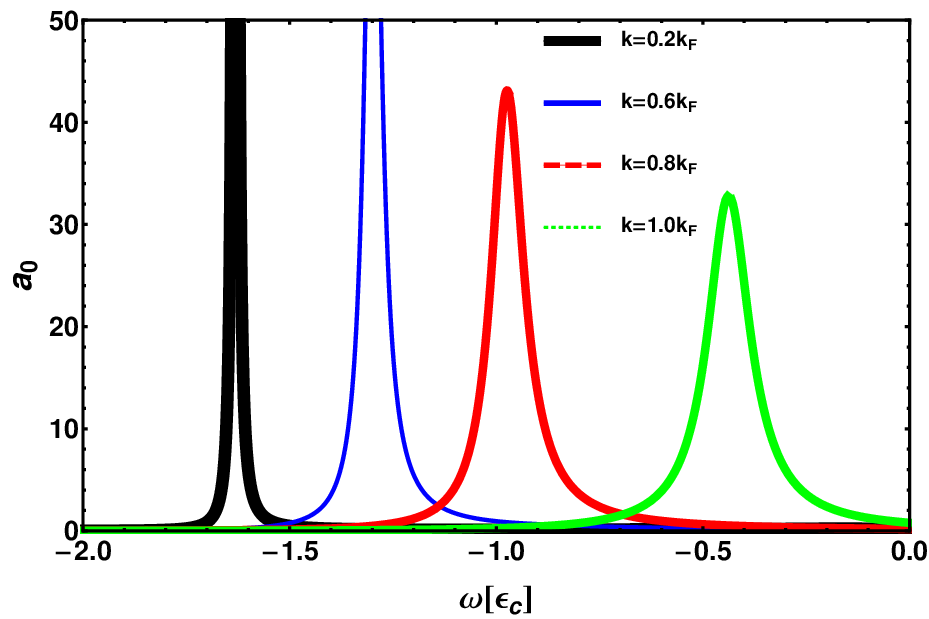}}
\centerline{\includegraphics[width=8cm]{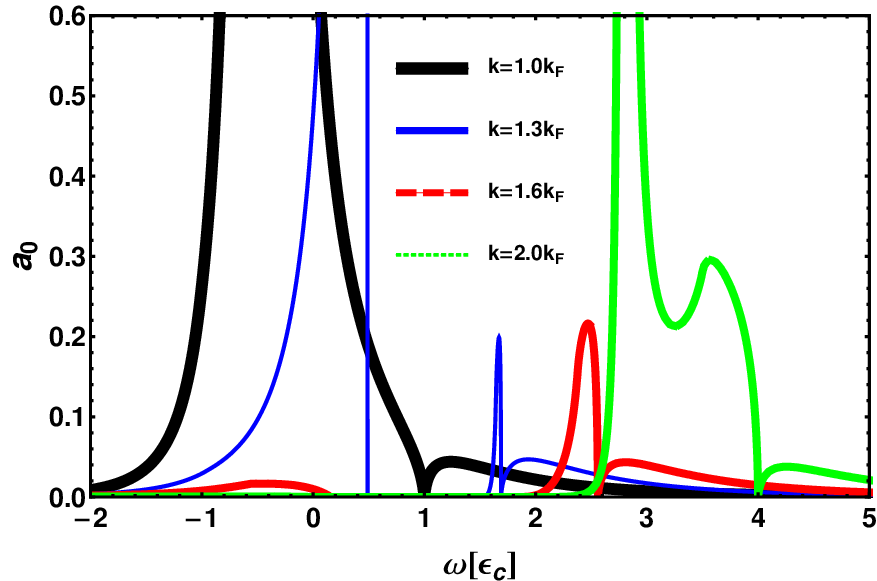}}
\caption{\label{a0} The (unrenormalized) non-selfconsistent electron spectral function (\ref{a}) for various momenta.}
\end{figure}

Next we calculate the real part of the selfenergy (\ref{Hilbert}) which
allows to discuss the spectral function of the electrons from the Dyson equation 
\be
a(k,\omega)=-2 {\rm Im}\left [\omega-{k^2\over 2 m}-\Sigma_F(k)-\Sigma^R(k,\omega)\right ]^{-1}.
\label{a}
\ee
In figure \ref{a0} we present this spectral function (\ref{a}) for various momenta. We see that the pole increases according to the expected dispersion $k^2/2m+\Sigma_F+\Sigma$ until the Fermi energy. Above, the spectral function shows quite a fragmented behaviour indicating that we have missed the correct pole. 

\begin{figure}
\centerline{\includegraphics[width=8cm]{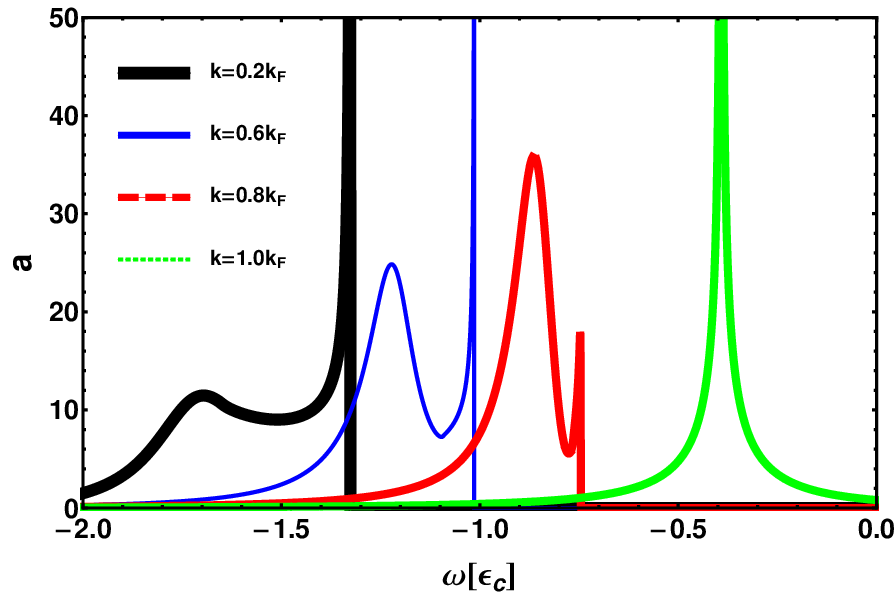}}
\centerline{\includegraphics[width=8cm]{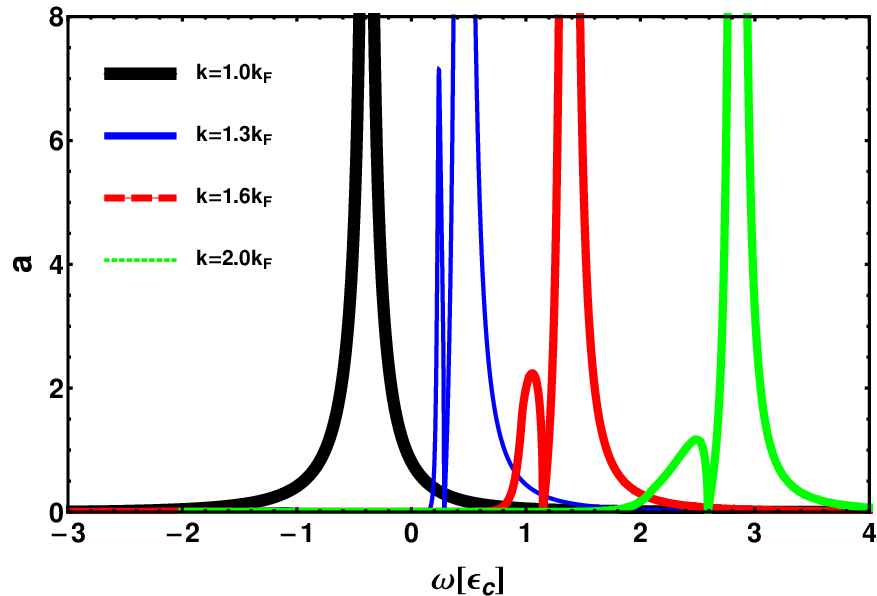}}
\caption{\label{as} The selfconsistent electron spectral function (\ref{a}) with $\Sigma(k,\omega+\Delta_k)$ for various momenta of figure \ref{a0}.}
\end{figure}

Moreover, there are two sum rules known for the spectral function, for derivation see \cite{M17b}. The first one, the norm conservation, is
\be
\int {d\omega\over 2 \pi} a(k,\omega)=1
\label{sumn}
\ee
and the second one, the energy-weighted sum rule, reads
\be
\int {d\omega\over 2 \pi} \omega a(k,\omega)={k^2\over 2 m}+\Sigma_F(k).
\label{sume}
\ee
Checking, one finds that below the Fermi momentum both sum rules are completed only within $5-10\%$ but with higher than Fermi momentum both are badly off. The reason by deeper inspection is that the energy argument of the selfenergy is not the energy $\omega$. In principle one has to meet the energy at the pole of the spectral function there. This creates a selfconsistency loop which has to be performed by iteration. As consequence this leads away from the $\omega$ argument of the perturbative $\Sigma(\omega)$ to a position $\omega+\Delta_k$. A very good short-cut is to consider this shift at the Fermi momentum but corrected by $\Delta_k \approx-\Sigma_F(k_F)-\Sigma(k_F,\epsilon_F)$. In fact this corrected form of spectral function towards selfconsistency completes both sum rules better that $0.01\%$ and are given in figure \ref{as}. The difference to figure \ref{a0} is visible. 

We see that below the Fermi momentum a sharp side peak develops which is vanishing at the Fermi momentum. Above this sharp side peak is suppressed again. Below zero a bound state pole is visible which vanishes for momenta around $2 k_f$ indicating that nesting is destroying the appearance of bound states.

\subsubsection{Extended quasiparticle spectral function}

According to the extended quasiparticle picture for the correlation function (\ref{gs}) we can also write the spectral function as
\be
a_{\rm EQP}(k,\omega)=G^>+G^<={2\pi \delta (\omega-\epsilon_k)\over 1-{\partial \Sigma(\omega)\over \partial \omega}}+{\Gamma(k,\omega)\over (\omega-\epsilon_k)^2}
\label{aeq}
\ee
with the dispersion $\epsilon_k=k^2/2m-\Sigma_F(k)-\Sigma(k,\epsilon_k)$.
This spectral function is the consistent expansion in second order potential according to (\ref{gs}).
The residue renormalizes the weight of the pole and the sum rules (\ref{sumn}) and (\ref{sume}) are completed \cite{M17b}.

\begin{figure}
\centerline{\includegraphics[width=8cm]{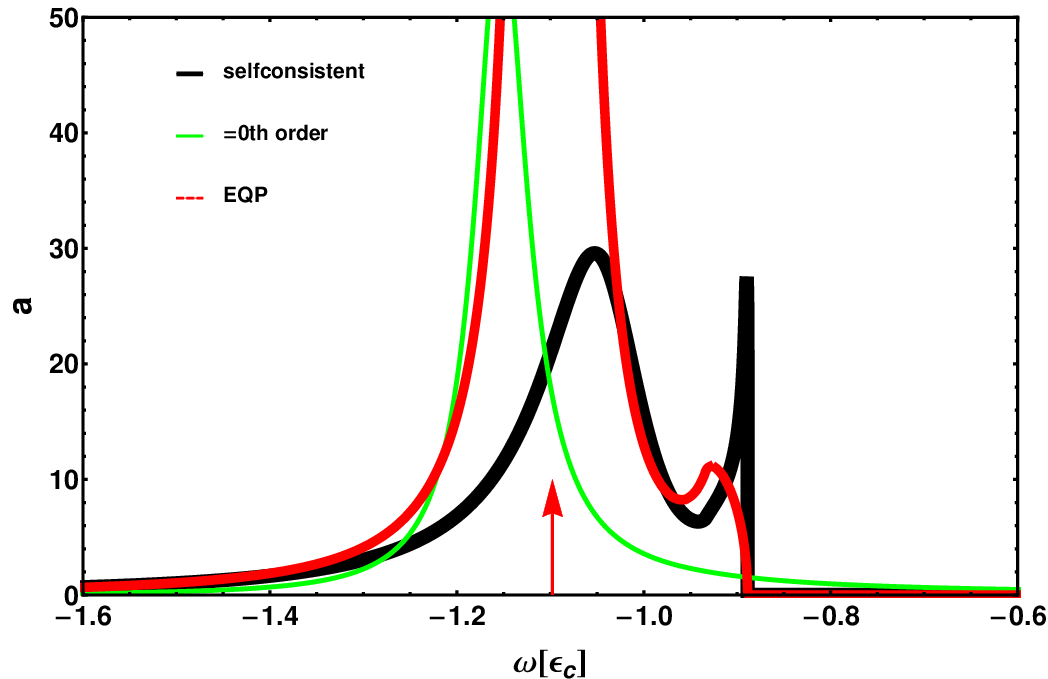}}
\centerline{\includegraphics[width=8cm]{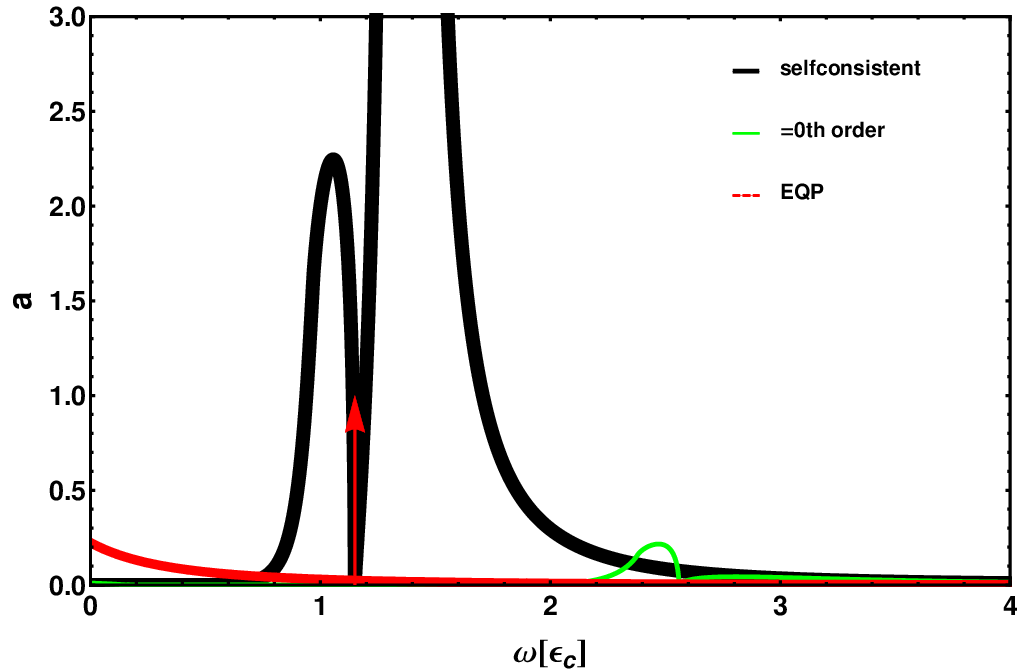}}
\caption{\label{aeqp} The selfconsistent electron spectral function (\ref{a}) with the non-selfconsistent as well as the extended quasiparticle one (\ref{aeq}
for momenta $k=0.7k_F$ above and $k=1.6k_F$ below. The arrow indicates the pole of the $\delta$-function.}
\end{figure}

In figure \ref{aeqp} we compare the selfconsistent spectral function with the extended quasiparticle one. One sees how the selfconsistent one approximates the correct pole which is indicated by an arrow and approaches the side band for higher momenta. 

\subsubsection{Quasiparticle energy and density of states}
\begin{figure}
\centerline{\includegraphics[width=8cm]{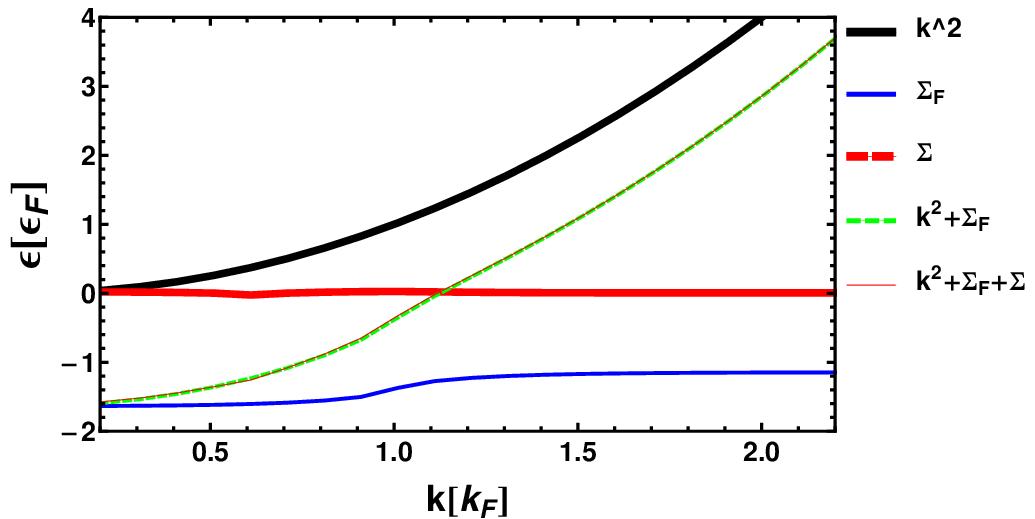}}
\centerline{\includegraphics[width=8cm]{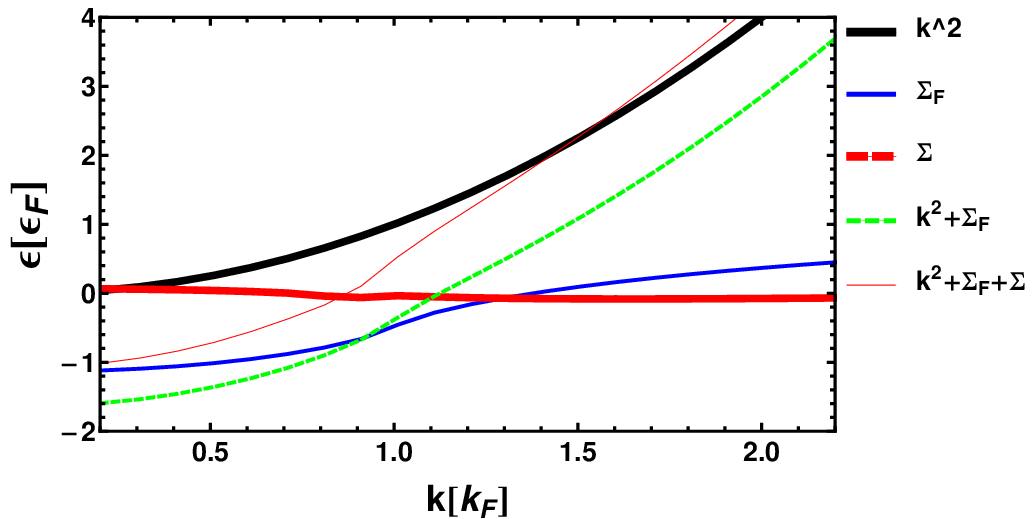}}
\centerline{\includegraphics[width=8cm]{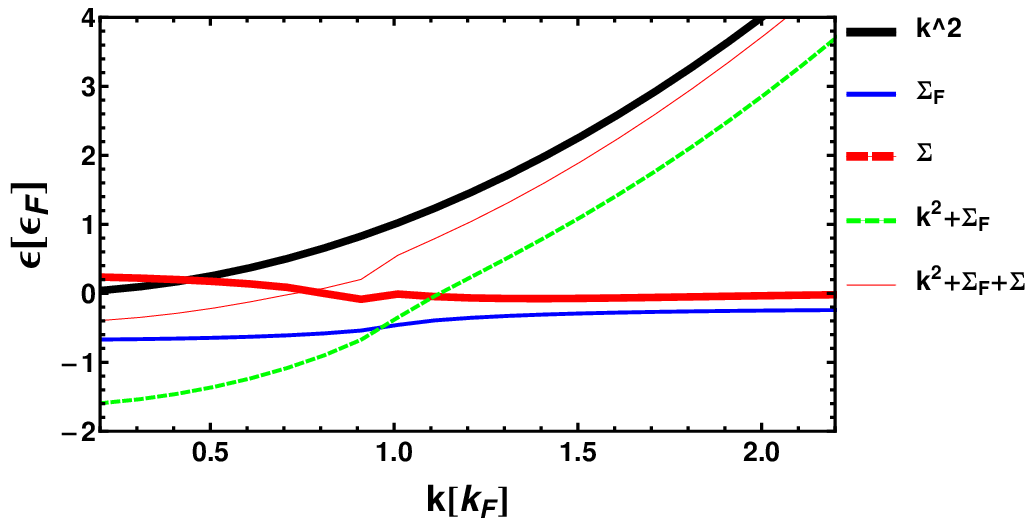}}
\caption{\label{energy} The quasiparticle energy dispersion (\ref{disp}) together with mean field and selfenergy contribution for $b=1, r_s=1$ (above), $b=0.1, r_s=1$ (middle) and $b=1, r_s=2$ (below).}
\end{figure}

The spectral function describes the one-particle excitations of the electrons. 
The quasiparticle excitation of the electrons are given by the main pole of the spectral function and according to the above discussion can be approximated by
\be
\epsilon_k={k^2\over 2 m}+\Sigma_F(k)+\Sigma(k,{k^2\over 2 m}+\Delta_k).
\label{disp}
\ee
In figure \ref{energy} we plot various contributions to the dispersion. If we compare the case of $b=1$ with the Coulomb limit $b=0.1$ we see that the first-order selfenergy becomes remarkable and compensates partially the strong meanfield contribution. Of course this is dependent on $r_s$. For illustrative purpose we plot also the case of $r_s=2$ seeing how the influence of the meanfield is further reduced.

\begin{figure}
\centerline{\includegraphics[width=8cm]{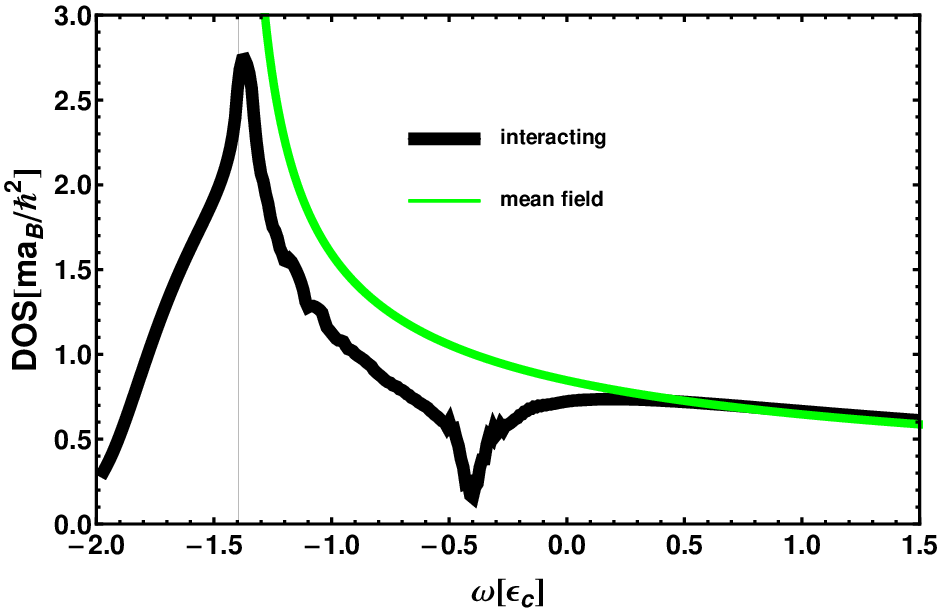}}
\caption{\label{DOS} The density of states (\ref{dos}) with the spectral functions (\ref{a}) of figures~\ref{a0} and \ref{as} (thick line) together with the mean field one $\sqrt{m/2(w-\Sigma_F(k_F))}$ (thin line) for $b=1, r_s=1$. The meanfield value of frequency is indicated by a vertical grid line.}
\end{figure}

From the spectral function we can also calculate the density of states
\be
{\rm DOS}(\omega)=\int{dk\over 2 \pi \hbar} a(k,\omega)
\label{dos}
\ee
which is plotted in figure \ref{DOS}. One sees how the meanfield density of states is approached at higher frequencies. At lower frequencies we get a reduction from the meanfield value showing no divergence. The shift of the bottom is nearly identical to the mean field value. The dip is the reminiscence of the gap in the excitation seen in the spectral functions in figures~\ref{a0} and \ref{as}.

\subsection{Structure factor and pair correlation function\label{SFCF}}

Now we are going to evaluate the response function (\ref{LCHI}) structure function (\ref{cw}) and the pair correlation (\ref{structfac}). 
Since we consider the first-order high-density expansion equivalent to the second-order expansion in the potential, we can expand Eq. (\ref{LCHI}) as
\be
\label{resHDE}
 \chi(q,\omega)&=&\Pi_{0}(q,\omega)+\lambda\; V(q)\Pi_{0}^2(q,\omega)\nonumber\\
 & &+\lambda\; \Pi_{se}(q,\omega)+\lambda \;\Pi_{ex}(q,\omega)
\ee
where we indicate the order of interaction by $\lambda$.
The first-order static structure factor (\ref{cw}) can be written according to (\ref{resHDE})
\be
S_1(x)=S_{V\Pi_0^2}(x)+S_{se}(x)+S_{ex}(x)
\ee
where we will use $x=q/2k_F$ in the following.
The analytical evaluation of $S_{V\Pi_0^2}$ and $S_{ex}$ for an infinitely-thin cylindrical wire can be found in \cite{VBMP17}. The contribution of the selfenergy to the structure factor  $S_{se}(q,\omega)$ turns out to be zero due to the $\omega$ integration. The sum of both corrections $S_{V\Pi_0^2}$ and $S_{ex}$ is given by \cite{VBMP19}
\ba
\label{SCyl1}
&S_1(x)\nonumber\\
&=\frac{g_s^2 r_s}{\pi ^2 x}\left \{
\begin{array}{ll}
\zeta(x) &, x<1
\cr
\zeta(x)-2 x \ln x \ln e^2 x&, x>1
\end{array}
\right . 
\end{align}
with
\be
\zeta(x)&=&(x+1)\ln (x+1)\ln \left ( {x^2 e^2\over x+1} \right )
\nonumber\\
&&+|x-1|\ln |x-1|\ln\left ({x^2 e^2\over |x-1|}\right ).
\label{z}
\ee


\subsection{Correlation energy}

Next we discuss the expression for the correlation energy per particle in second-order perturbation theory \cite{LL79}, i.e. second Born approximation with exchange (\ref{ecorr}).
%
%
For contact potentials we have to subtract an infinite value $\sim n_{p_1}n_{p_2}$ in order to reach convergence  which is a renormalization of contact potential. For finite-range potentials we have an intrinsic cut-off due
to the range of interaction and such problem does not occur as we see a posteriori.

\begin{figure}
\includegraphics[width=9cm]{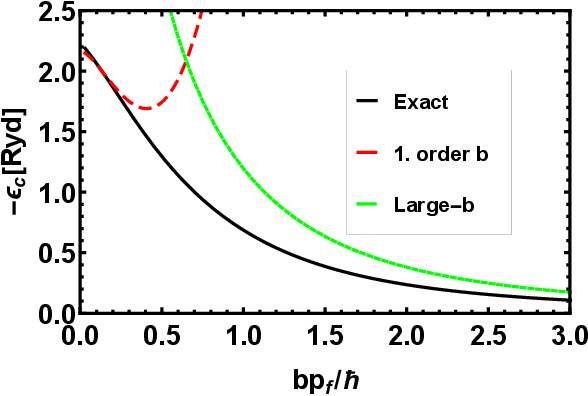}
\caption{\label{ecorr23_2}The correlation energy per particle (\ref{ec1}) together with the first-order analytical expansion (\ref{Int_Corr}) as well as the large-b expansion.}
\end{figure}

We scale all momenta again by
the Fermi momentum $k_{\uparrow\downarrow}=\pi \hbar n/g_s$ as $p_1=k/k_{\uparrow\downarrow}$, $p_2=p/k_{\uparrow\downarrow}$, and $x=q/2k_f$. The occupation factors restrict the integration range that from $1>p_1^2$, $1>p_2^2$, $(p_1+2 x)^2>1$ and $(p_2+2 x)^2>1$ follows the two cases $0<x<1$ with $1-2x<p_1<1, -1<p_2<2 x-1$ and $1<x$ with $-1,p_1,p_2<1$. Presenting the energy in terms of ${\rm Ryd} =e^2/4 \pi \epsilon_0a_B$ as $\epsilon_c=E_c/n/{\rm Ryd}$ we obtain 
\ba
{\epsilon_c}=&-{1\over 4 \pi^2} \left [\int\limits_{0}^1 {dx\over x}\, \Lambda_x^<+\int\limits_{1}^\infty {dx\over x}\, \Lambda_x^>\right ]
\label{ec1}
\end{align}
with $b=\bar b 2k_f$.
The $p_1$ and $p_2$ integrations can be carried out analytically and yield
for $x>1$
\ba
&\Lambda_x^>=\!\int\limits_{-1}^1 \!\!dp_1\! \int\limits_{-1}^{1} \!\!dp_2 {v(2k_f x)[v(2k_f x)\!-\!v(k_f p_1\!-\!k_fp_2\!+\!2 k_f x)]\over 2 x(p_1\!-\!p_2\!+\!2 x)}
\nonumber\\
&=\frac{K_0(2 b x)}{b} \left\{
G_{2,4}^{3,1}\left({bx}-b,\frac{1}{2}|
\begin{array}{c}
 1,\frac{3}{2} \\
 \frac{1}{2},\frac{1}{2},\frac{1}{2},0 \\
\end{array}
\right)
\right .
\nonumber\\&
\left .
-2 G_{2,4}^{3,1}\!\left(\!{b x},\frac{1}{2}|
\begin{array}{c}
 1,\frac{3}{2} \\
 \frac{1}{2},\frac{1}{2},\frac{1}{2},0 \\
\end{array}
\!\!\right)\!+\!G_{2,4}^{3,1}\left({bx} \!+\!b,\frac{1}{2}|
\begin{array}{c}
 1,\frac{3}{2} \\
 \frac{1}{2},\frac{1}{2},\frac{1}{2},0 \\
\end{array}
\!\!\right)
\right .
\nonumber\\&
+4 b K_0(2b x)\left [2 x \ln \left ({x^2\!-\!1\over x^2}\right ) \!+\!\ln \left ({x\!+\!1\over x\!-\!1}\right )^2\right ]
 \biggr \}
\end{align}
and for $0<x<1$
\ba
&\lambda_x^<=\!\!\!\int\limits_{1\!-\!2x}^1 \!\!\!\!dp_1\!\!\! \int\limits_{-1}^{2x\!-\!1}\!\!\!\!dp_2 {v(2 k_f x)[v(2 k_f x)\!-\!v(k_f p_1\!-\!k_f p_2\!+\!2 k_f x)]\over 2 x(p_1\!-\!p_2\!+\!2 x)}
\nonumber\\
&=\frac{K_0(2 b x)}{b} \left\{
G_{2,4}^{3,1}\left(\!b\!+\!{bx}),\frac{1}{2}|
\begin{array}{c}
 1,\frac{3}{2} \\
 \frac{1}{2},\frac{1}{2},\frac{1}{2},0 \\
\end{array}
\!\right)
\right .
\nonumber\\&
\left .+G_{2,4}^{3,1}\left(\!b\!-\!{bx}),\frac{1}{2}|
\begin{array}{c}
 1,\frac{3}{2} \\
 \frac{1}{2},\frac{1}{2},\frac{1}{2},0 \\
\end{array}
\right)-2 G_{2,4}^{3,1}\left(2x,\frac{1}{2}|
\begin{array}{c}
 1,\frac{3}{2} \\
 \frac{1}{2},\frac{1}{2},\frac{1}{2},0 \\
\end{array}
\!\right)
\right.
\nonumber\\&
+4 b K_0(2 b x)\left(2 \ln \left(1-x^2\right)+2 x \ln
   \left(\frac{x+1}{1-x}\right)\right)
\biggr \}
\end{align}
with the the Meijer G function. The last x-integral can be done numerically and the result is seen in figure~\ref{ecorr23_2}. It shows that the ground state correlation energy is decreasing continuously with increasing width. This means that the one-dimensional systems is unstable compared to the two-dimensional system which is the large-b limit presented as well in figure \ref{ecorr23_2}. We see how the exact expression (\ref{ec1}) interpolates between both limits.

\subsubsection{Small-b expansion}

We have two ways to calculate the correlation energy via the charging formula (\ref{ef}) and via the selfenergy (\ref{Ec}). The Fock or exchange term has been given already by (\ref{eFx}) and was shown to yield equivalent results with the calculation by the selfenergy (\ref{fockproof}). Lets check this with the small-b expansion. 

First we give the result via the charging formula.
The correlation energy per particle in Eq. (\ref{ec}) in the small-$b$ limit for a cylindrical wire is given by \cite{VBMP19}
\be
\label{Int_Corr}
\epsilon_c
&=&\frac{1}{4r_s}\bigg\{\Lambda_{(x<1)}+\Lambda_{(x>1)} \bigg\}
\ee
where we use the small $b$ expansion of $v(x)$. The result for $x<1$ is
\ba
\label{lamda_clyl1}
&\Lambda_{(x<1)}=\int^1_0 v(x) [S_1(x)]_{x<1}\;dx
\nonumber\\
&=\frac{r_s g_s^2}{12\pi^2} \bigg\{42 \zeta (3) \ln \left(\frac{bk_F}{8}\right)
+48 (\ln
   (2)-2) \ln (2) \ln (bk_F)
\nonumber\\
&+48 \left(-2
   \text{Li}_4\left(\frac{1}{2}\right)+\ln ^2(2)+\gamma  \left(\ln
   ^2(2)-\ln (4)\right)+\ln (4)\right)\nonumber\\
   & +42 (\gamma -1) \zeta (3)+\pi
   ^4-4 \log ^3(2) (12+\ln (2))+4 \pi ^2 \ln ^2(2)\bigg\},
\end{align}
and for $x>1$ it is
\ba
\label{lamda_clyg1}
&\Lambda_{(x>1)}=\int^\infty_1 v(x) [S_1^{Cy.}(x)]_{x>1}\; dx
\nonumber\\
&=-\frac{2r_s g_s^2}{\pi^2} \bigg\{\frac{7}{4} \zeta (3) \left(\ln
   \left(\frac{bk_F}{8}\right)+\gamma -1\right)-4\text{Li}_4\left(\frac{1}{2}\right)
\nonumber\\
& +\frac{17 \pi
   ^4}{360}+(\ln (2)-2) \ln (4) \ln
   (bk_F)-\frac{\ln ^4(2)}{6}-2 \ln ^3(2)\nonumber\\
   & +\frac{1}{6} \pi ^2 \ln
   ^2(2)+2 \gamma  \ln ^2(2)+2 \ln ^2(2)+\ln (16)-4 \gamma  \ln
   (2)\bigg\},
\end{align}
where $\zeta(s)$ is the Riemann zeta function and $\text{Li}_n(z)$ is the polylogarithm function \cite{a84}. Adding Eq.(\ref{lamda_clyl1}) and (\ref{lamda_clyg1}), major cancellations occur and one obtains the known correlation energy as
\be
\label{exact}
\epsilon_c(r_s)=-\frac{\pi^2}{360}
\ee
which is the result of the conventional perturbation theory\cite{Loos13,LG16}
and in excellent agreement with variational quantum Monte Carlo simulation \cite{Vinod18c}.

As comparison we now calculate the small-b expansion via the second Born approximation (\ref{ec1}). We use first the lowest-order of (\ref{v})
\ba
&v(2 k_f x)[v(2 k_f x)-v(k_f p_1-k_f p_2+2 k_f x)]
\nonumber\\&
=4 (\gamma+\ln {b/2}+\ln{2 x}) [\ln 2x-\ln (p_1-p_2+2 x)]+o(b^2).
\label{vs}
\end{align}
The $p_1$ and $p_2$ integrals read
\ba
&\Lambda_x^>=4 (\gamma+\ln {b/2}+\ln 2x)\left [\xi(x)
-2 x \ln x\ln e^2 x
\right ].
\end{align}
For $0<x<1$ one gets
\ba
&\Lambda_x^<=4 (\gamma+\ln {b/2}+\ln (2x))\xi(x).
\end{align}
Comparing with (\ref{SCyl1}) we see exactly the same expressions $\xi(x)$. This means that in the static perturbation theory the structure factor is silently contained but not possible to identify directly here.

Integrating further we obtain
\be
&&\int\limits_1^\infty {dx\over x} \Lambda_x^>=
(\gamma+\ln {b/2}) [8\ln^2(2)-16 \ln 2+7 \zeta(3)]
\nonumber\\&&
-16 \text{Li}_4\left(\frac{1}{2}\right)-7 \zeta (3) [1+\ln 2]+\frac{17
   \pi ^4}{90}+\frac{2}{3} \pi ^2 \ln ^2 2
\nonumber\\&&
+16 \ln 2-\frac{2}{3} (\ln
   2-6)^2 \ln^2 2
\label{l1}
\ee
and
\be
&&\int\limits_0^1 {dx\over x} \Lambda_x^<=
-(\gamma+\ln {b/2}) [8\ln^2(2)+16 \ln 2+7 \zeta(3)]
\nonumber\\&&
16 \text{Li}_4\left(\frac{1}{2}\right)+7 \zeta (3)
   (1+\ln 2)-{\pi^4\over 6}-{2 \pi^2\over 3} \ln ^2 2
\nonumber\\&&
\!-\!16 \ln
   2\!+\!{2\over 3}(\ln 2\!-\!6)^2 \ln^2 2\biggr ].
\label{l2}
\ee
Adding (\ref{l1}) and (\ref{l2}) we get
\be
\epsilon_c=-{\pi^2\over 360}
\ee
which is exactly (\ref{exact}). So both ways, the charging energy formula and the selfenergy gives the same results.

With the same means we can calculate the next term of (\ref{v}).
We obtain the next order in $b$
\be
&&\int\limits_1^\infty {dx\over x} \Lambda_x^>=
(\gamma+\ln {b/2})^2 {16\over 3}(1-\ln 2)
\nonumber\\&&
+(\gamma+\ln {b/2}) {2\over 9}[3\pi^2-68-4\ln 2(9 \ln 2-29)]
\nonumber\\&&
+{1\over 54}[452+(36 \pi^2-1448)\ln 2-48\ln^2 2(3\ln 2-19)
\nonumber\\&&
-30\pi^2+315 \zeta(3)]
\label{lb1}
\ee
and
\be
&&\int\limits_0^1 {dx\over x} \Lambda_x^<=
-(\gamma+\ln {b/2})^2 {16\over 3}(1-\ln 2)
\nonumber\\&&
+(\gamma+\ln {b/2}) {2\over 9}[-\pi^2+74+4\ln 2(9 \ln 2-29)]
\nonumber\\&&
-{1\over 54}[656+(12 \pi^2-1510)\ln 2-48\ln^2 2(3\ln 2-19)
\nonumber\\&&
+10\pi^2+27 \zeta(3)]
\label{lb2}
\ee

 which results into
\be
&&\epsilon_c=-\frac{\pi ^2}{360}
\nonumber\\
&&
\!-\frac{b^2 \left(6 \left(3\!+\!\pi ^2\right) \ln (b)\!+\!72 \zeta (3)\!+\!6 \gamma
   \left(3\!+\!\pi ^2\right)\!-\!5 \pi ^2\!-\!51\right)}{108 \pi^2}.
\nonumber\\
&&
\ee

\subsection{Reduced density matrix\label{reddd}}

\begin{figure}
\includegraphics[width=9cm]{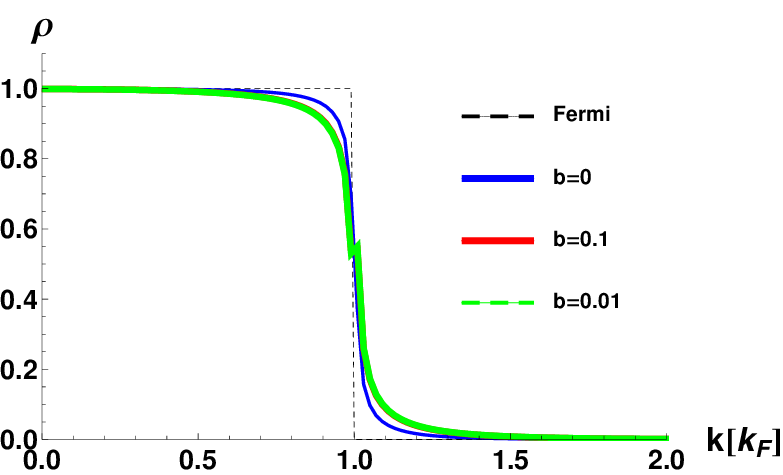}
\caption{\label{reduced} The reduced density matrix (\ref{red}) for various width  parameter $b$ together with the Fermi function and artificially $4 r_s^2/\pi^4=1$.}
\end{figure}

Finally we calculate the reduced density matrix (\ref{red}) or explicitly (\ref{rho}). The integration ranges due to the occupation factors are worked out in appendix~\ref{ared}.
For small $b$ parameter this can be evaluated analytically. One finds that at $k=1 k_F$ the reduced density matrix has a singularity in that it diverges of both sides with opposite sign.
The result for the numerical integration over the selfenergies (\ref{red}) are given in figure \ref{reduced} for various $b$ parameter. One sees that in the Coulomb limit the divergence is seen as a small wiggle around $0.5$. 

This divergence has been discussed in \cite{M23} and a Pad\'e regularization was suggested. It consists of the extended quasiparticle approximation used so far and an additional expansion around the Fermi energy both interpolated by a function rapidly vanishing outside the Fermi energy. We can therefore assume such regularization and it would subtract the divergence on both sides which has here the form $c_1\ln (k-1)+c_2\ln(k-1)^2+c_3 \ln (1-k)^3$.
One obtains the interesting limiting law for the harmonic potential (\ref{v})
\be
\lim\limits_{k\to 1\pm 0}\rho_k=\left \{
\begin{array}{l}
{1\over 12} [3 + \ln(2)^3]\approx 0.277752\cr
1+{1\over 12} [-3 + \ln(2)^3]\approx 1-0.222248
\end{array}
\right .
\ee
which exactly approaches the jump 
\be
\Delta \rho=\frac 1 2
\label{jump}
\ee 
at the Fermi momentum. Since we know that the momentum distribution is finite at the Fermi momentum, the interpolation between expansion at the Fermi energy and the extended quasiparticle approximation easily accounts for this finite jump subtracting not only the divergent terms on both sides but also the jump \cite{M23}. The corresponding analytical expression for the momentum distribution is somewhat lengthy but trivially obtained by the formulas in appendix~\ref{ared}. 

It is instructive to consider the limit of contact potentials $v(q)=1$. Then one obtains analytically
\ba
&\rho_k=1+{r_s^2(g_s-1)\over \pi^4}
\nonumber\\
&\times\!
\left \{\!
\begin{array}{cc}
{\ln(1\!+\!k)\over (1\!-\!k)^2}\!+\!{\ln(1\!-\!k)\over (1\!+\!k)^2} \!-\!2(1\!+\!2\ln 2){1\!+\!k^2\over (k^2\!-\!1)^2}& 0<k<1
\cr
&
\cr
{1+2 \ln 2-2\ln(k-1)\over (1+k)^2}&1<k<3
\cr
&
\cr
{4\over k^2-1)^2}&k>3
\end{array}
\right .
\end{align}
and one sees that near the Fermi momentum $k=1\pm \eta$ we have from both sides
\be
\rho_k\approx\left \{
\begin{array}{cc}
1+{1-2\ln 2\over 4}+\frac 1 2 \ln \eta & k=1-\eta
\cr
{1+2\ln 2\over 4}-\frac 1 2 \ln \eta & k=1+\eta
\end{array}
\right .+o(\eta).
\ee
Again due to Pad\'e expansion we subtract a regularizing term $\rho_k=1+(\rho^>_k-\rho^r_k)-(\rho^<_k-\rho^r_k)$ to get rid of divergences and see that the jump at the Fermi momentum approaches a smaller value than (\ref{jump}) of 
\be
\Delta \rho=1-\ln 2\approx 0.3068.
\ee
Again this can be included additionally into the Pad\'e term rendering the value of the momentum distribution unique at the Fermi energy.

\section{Summary\label{summ}}

We have presented two approaches to the correlation energy, one by the structure factor with the pair correlation function and one by the Dyson equation with the selfenergy. Both are rooted in approximating the two-particle Green function appropriately. Different resulting forms are compared and it is shown how they coincide if the same level of approximations is used. The equivalence is obtained within the extended quasiparticle picture for the single-particle propagators and self energies. 

For a one-dimensional quantum wire of Fermions the approximations are illustrated and the self energies are explicitly discussed. A gap appears which results into a splitting of excitation lines in the spectral function of holons and antiholons. Also bound states are visible destroyed by higher momentum around nesting. The meanfield leads to an effective mass which shows the onset of Peierls-like transition at twice the Fermi energy. The density of states in Born approximation and meanfield are compared and the correlation effects are identified. The width dependence of the correlation energy is calculated and compared with the analytical results of small and large width expansions. The momentum distribution shows a divergence in approaching the left and right side of the Fermi energy. The occurring divergences and jumps at the Fermi energy are subtracted due to a regularization scheme of Pad\'e which interpolates between the extended quasiparticle approximation and an expansion at the Fermi energy.

\begin{acknowledgments}
This paper is devoted to the memory of Paul Ziesche (1933-2022) with whom the authors had many enlightening discussions.
K.M.\ acknowledges support from DFG-project
MO 621/28-1. K.N.P. acknowledges the grant of honorary senior scientist position by National Academy of Sciences of India (NASI) Prayagraj. V.A. acknowledge the support in the form of SERB-Core research Grant No. CRG/2023/001573.
\end{acknowledgments}

\appendix
\section{Integration range for the selfenergy (\ref{self1})\label{range}}

Performing the restrictions of the $\Theta$ functions the following integration range for $\Sigma^<$ appears. It is only non-zero for $W=k^2-\omega>0$ and
for $0\le k<1$
\ba
& p\!>\!1,W\!<\!2(k\!-\!1)^2:{\rm Max}(1,k\!+\!\sqrt{2W})\!<\!p\!<\!1\!+\!{W\over 2(1\!-\!k)}
\nonumber\\
&p\!<\!\!-\!1,W\!<\!2(1\!+\!k)^2: \!-\!1\!-\!{W\over 2(1\!+\!k)}\!<\!p\!<\!{\rm Min}(\!-\!1,k\!-\!\sqrt{2W})
\end{align}
and for 
$k\!>\!1$
\ba
&p\!<\!-\!1,2 k^2\!<\!W\!<\!2 (1\!+\!k)^2:
\nonumber\\
&\qquad \!-\!1\!-\!{W\over 2(1\!+\!k)}\!<\!p\!<\!{\rm Min}(\!-\!1,k\!-\!\sqrt{2W})
\nonumber\\
&p\!<\!-\!k\!<\!-\!1: \!-\!1\!-\!{W\over 2(1\!+\!k)}\!<\!p\!<\!-\!1
\nonumber\\
&\!-\!k\!<\!p\!<\!-\!1, 2(k\!-\!1)^2\!<\!W\!<\!2 (k^2\!-\!1), k\!>\!\frac 5 3: 
\nonumber\\
&\qquad {\rm Max}(\!-\!k,1\!-\!{W\over 2(k\!-\!1)}\!<\!p\!<\!{\rm Min}(\!-\!1,k\!-\!\sqrt{2W})
\nonumber\\
&\qquad \qquad k\le \frac 5 3: {\rm Max}(\!-\!k,1\!-\!{W\over 2(k\!-\!1)})\!<\!p\!<\!-\!1
\nonumber\\
&2(k^2\!-\!1)\!<\!W\!<\!2 k^2:\!-\!k\!<\!p\!<\!{\rm Min}(\!-\!1,k\!-\!\sqrt{2W}).
\end{align}

The integration range for $\Sigma^>$ is somewhat simpler.
for $0\le k<1$
\ba
{\rm Max}(-1,1\!+\!{W\over 2(1\!-\!k)})\!<\!p\!<\!{\rm Min}(1,\!-\!1\!-\!{W\over 2(1\!+\!k)})
\end{align}
and for 
$k\!>\!1$ and $W>0$
\ba
&k\ge 3 \,{\rm or}\, (k<3 \,{\rm and}\, 2(k-1)^2>W):
\nonumber\\
&\qquad
{\rm Max}(-1,1\!+\!{W\over 2(1\!-\!k)})\!<\!p\!<\!{\rm Min}(1,k\!-\!\sqrt{2W})
\nonumber\\
&1<k< 3 :
-1<\!p\!<\!{\rm Min}(1,\!-\!1\!-\!{W\over 2(1\!-\!k)})
.
\end{align}

\section{Integration range for the reduced density matrix\label{ared}}
We scale all momenta by the Fermi momentum to obtain
\be
\rho_k&&=n_k+{4 r_s^2\over \pi^4}\iint dp dq {V_q(g_s V_q-V_{p-k-q})
\over 
[2 q(k-p+q)]^2}
\nonumber\\
&&\times
\left \{
[k>1][p^2>1][1>(k+q)^2][1>(p-q)^2]
\right .
\nonumber\\
&&\left .-[1>k][1>p^2][(k+q)^2>1][(p-q)^2>1] \right \}
\ee
where we can restrict to positive $k$ since $\rho_{-k}=\rho_{k}$ which one sees by interchanging sign of $p,q$. The first part appears for momenta larger than Fermi momentum, $k>1$ and the second part for $0<k<1$. Discussing the
integration range for $0<k<1$ we have $-1<p<1$ and $q<p-1$ or $q>p+1$ as well as $q<-1-k$ or $q>1-k$. This provides two cases
\ba
&(a):\, -1\!<\!-k\!<\!p\!<\!1:\, -\infty\!<\!q<-1\!-\!k\, {\rm or}\,  p\!+\!1\!<\!q\!<\!\infty
\nonumber\\
&(b):\, -k>p>-1:\, -\infty<q<p-1\, {\rm or}\,  1-k<q<\infty.
\end{align}
Together this provides the integration range
\be
&&\int\limits_{-1}^{-k} \!\! dp \left [ \int\limits_{-\infty}^{p-1} \!\! dq+\int\limits^{\infty}_{1-k} \!\! dq\right ]
+
\int\limits^{1}_{-k} \!\! dp \left [ \int\limits_{-\infty}^{-k-1} \!\! dq+\int\limits^{\infty}_{p+1} \!\! dq\right ]
\nonumber\\
&&=
\int \limits_2^\infty d q \int \limits_{-1}^1 \!\! dp+\int \limits_{1-k}^2 d q \int \limits_{-1}^{q-1} \!\! dp+ (k\leftrightarrow -k).
\ee

For $k>1$ we have $p>1$ or $p<-1$ and two conditions for $q$
\be
&&p-1<q<p+1, \qquad -1-k<q<1-k.
\label{ranged}
\ee
For $p>1$ we have $1-k<0<p-1$ and there is no common overlap for the range of $q$. Since both ranges (\ref{ranged}) have the length of $2$ we have to cases of finite overlap
\be
(1):&&p-1<q<1-k
\nonumber\\
&&{\rm if}\, -1-k<p-1<1-k<p+1<0\nonumber\\
(2):&&-k-1<q<p+1
\nonumber\\
&&{\rm if}\, p-1<-k-1<p+1<1-k<0.
\ee
Case (1) translates into $-k<p<Min(-1,2-k)$ which divides into two cases
\ba
(a):\, 1\!<\!k\!<\!3,\,-k\!<\!p\!<\!-1;\,(b):\,k\!>\!3,\, -k\!<\!p\!<\!2\!-\!k.
\end{align}
Case (2) yields $-2-k<p<-k$.
Combining case (1) and (2) we find
for $1\le k<3$
\ba
&\int\limits_{-2-k}^{-k} \!\!\!\! dp \int\limits_{-1-k}^{p+1} \!\!\!\! dq+
\int \limits_{-k}^{-1}\!\! d p \int \limits_{p-1}^{1-k} \!\! dq
=\int \limits^{-2}_{k-1} \!\!d q \int \limits_{q-1}^{q+1} \!\! dp
+\int \limits_{-1-k}^{1-k} \!\!\!\!d q \int \limits_{q-1}^{-1} \!\!\!\! dp
\end{align}
and for $k>3$
\be
\int\limits_{-2-k}^{-k} \!\!\! dp \int\limits_{-1-k}^{p+1} \!\!\!\! dq+
\int \limits_{-k}^{2-k} \!\!\!d p \int \limits_{p-1}^{1-k} \!\! dq=
\int \limits_{-1-k}^{1-k} \!\!\!d q \int \limits_{q-1}^{q+1} \!\! dp
.
\ee




%

\end{document}